\documentclass[a4paper,12pt]{article}

\usepackage{amsmath, amssymb, graphicx, physics, bm}
\usepackage[a4paper, margin=1in]{geometry}
\usepackage{tikz}
\usetikzlibrary{arrows.meta,calc}
\usepackage{jheppub, appendix, mathrsfs} 
\usepackage{slashed}

\title{\boldmath Spacetime Duality Beyond Conformality}

\author{Jeff Murugan$^{a}$ and Horatiu Nastase$^{b}$}
\affiliation{$^{a}$The Laboratory for Quantum Gravity \& Strings,\\
Department of Mathematics and Applied Mathematics,\\
University of Cape Town, Private Bag, Rondebosch, 7701,\\
South Africa\\
$^{b}$Instituto de Física Téorica,\\ UNESP-Universidade Estadual Paulista\\
R. Dr. Bento T. Ferraz 271, Bl. II, Sao Paulo 01140-070, SP, Brazil}

\emailAdd{jeff.murugan@uct.ac.za}
\emailAdd{horatiu.nastase@unesp.br}

\abstract{We extend the spacetime duality programme of Burgess \textit{et.al.} to massive theories in 1+1 dimensions. For the massive scalar, a heat-kernel computation tracking three contributions to the conformal-mode effective action reveals that the naive leading correction $\sim m^2(e^{\phi} -1)$ to the Liouville action cancels exactly, with the genuine leading deformation being $-\frac{m^2}{16\pi}(e^{\phi}-1)^{2}$. This breaks self-duality and renders the dual theory for the Lagrange multiplier field $\Lambda$ non-local. For the massive Dirac fermion, two independent derivations establish that the fermion mass dresses under conformal scaling as $m \to m\, e^{\phi/2}$, reflecting the Weyl weight $\frac{1}{2}$ of the two-dimensional spinor. Via the Coleman-Mandelstam bosonisation, this transfers to the mass bilinear as $\mu\cos(\beta\vartheta) \to \mu e^{\phi/2}\cos(\beta\vartheta)$, producing a coupled Liouville-sine-Gordon system as the natural starting point for the fermionic construction. Both results are interpreted in terms of the determinant line bundle over Met($\Sigma$)/Diff($\Sigma$).
}

\begin{document}
\maketitle
\flushbottom

%%%%%%%%%%%%%%%%%%%%%%%%%%%%%%%%%%
\section{Introduction}
\label{sec:intro}

The standard duality algorithm \cite{Buscher1987, RocekVerlinde1992} takes a
field theory $\mathscr{F}$ with a global internal symmetry $G$, gauges $G$ by
introducing a dynamical gauge field $A$, and imposes the flatness constraint
$F = \mathrm{d}A = 0$ via a Lagrange multiplier field $\Lambda$. Integrating
over $\Lambda$ and fixing a gauge for $G$ recovers $\mathscr{F}$; integrating
instead over the original matter fields and $A$ yields the dual theory
$\widetilde{\mathscr{F}}$, with $\Lambda$ as the fundamental dual variable. The
paradigmatic example is of course scalar T-duality, where $G$ is the shift symmetry
$X \to X + \omega$, but the same algorithm also underlies bosonization in two
dimensions \cite{BQ1994}, the equivalence of antisymmetric tensor fields of
complementary rank, and much of the modern duality web in
string theory and quantum field theory \cite{Seiberg2016, Karch:2016sxi, Murugan:2016zal}.\\

\noindent
In \cite{BQ1998} Burgess \textit{et.al.} proposed a natural
and intriguing generalisation of the duality algorithm; replace the internal symmetry $G$ by a \emph{spacetime}
symmetry. Gauging Lorentz invariance amounts to coupling the matter theory to a
dynamical metric $g_{\mu\nu}$; the constraint that sets $g_{\mu\nu}$ equal to a
fixed background $h_{\mu\nu}$ is then implemented by a scalar Lagrange
multiplier $\Lambda$ coupled to the difference of curvature scalars. Exchanging
the order of integration in the resulting master path integral yields a dual
theory whose fundamental field is $\Lambda$. Applied to a massless scalar in
$1+1$ dimensions, this procedure returns the same theory \textit{i.e.} the scalar is
self-dual under spacetime duality. Applied to a massless Dirac fermion on the other hand,
produces a free scalar, in essence recovering bosonization in a gravitational guise and
offering an alternative to the standard current-algebra derivation
\cite{BQ1994}. For $(2,2)$-supersymmetric models, the construction is more
powerful with the spacetime duality interchanging chiral and twisted-chiral multiplets,
making contact with mirror symmetry \cite{BQ1998}.\\

\noindent
Despite these successes, the original paper \cite{BQ1998} restricted attention
to massless, conformally invariant theories. This is not merely a technical
convenience. The entire mechanism, the Liouville action emerging from the
conformal anomaly, the Gaussian integral over the conformal mode, and the exact
cancellation of functional determinants, relies on the matter theory being a
CFT, whose path-integral measure is a section of the determinant line bundle
over $\mathrm{Met}(\Sigma)/\mathrm{Diff}(\Sigma)$ with curvature controlled by
the central charge. Any mass term breaks conformal invariance, and it is not obvious whether spacetime duality survives, and if so in what form. In this paper we answer this question. Our main results are as follows:

\begin{itemize}
\item For the \textbf{massive scalar}, a careful heat-kernel computation
tracking three distinct contributions to the conformal-mode effective action namely, the path-integral measure anomaly, the matter determinant, and the
$\Delta_{LM}$ constraint prefactor, reveals that the naive leading
correction $m^2(e^\phi - 1)$ cancels exactly. The genuine leading
correction to the Liouville action is $-\frac{m^2}{16\pi}(e^\phi-1)^2$,
quadratic in the conformal-factor deviation, which renders the conformal-mode
path integral non-Gaussian and breaks self-duality. The dual theory for
$\Lambda$ is non-local, with a kinetic operator interpolating between
the massless free-scalar result in the ultraviolet and a strongly suppressed
gapped theory in the infrared; it reduces to the massless result of \cite{BQ1998}
as $m \to 0$. For the interacting scalar (with quartic self-coupling), integrating
out the Hubbard--Stratonovich auxiliary field produces a logarithmic potential
for $\Lambda$ rather than a sine-Gordon potential.

\item For the \textbf{massive Dirac fermion}, the essential new ingredient is
the gravitational dressing of the fermion mass. Two independent derivations confirm that under the conformal rescaling $g = e^\phi h$, the
fermion mass dresses as $m \to m\,e^{\phi/2}$, not $m\,e^{\phi/4}$ as one
might naively expect from the spinor field redefinition factor. The exponent
$\phi/2$ reflects the Weyl weight $1/2$ of the two-dimensional Dirac spinor
and distinguishes the fermionic construction sharply from the scalar case. Via
the standard Coleman--Mandelstam bosonisation \cite{Coleman1975,
Mandelstam1975}, this dressing transfers to the mass bilinear as
$\mu\cos(\beta\vartheta) \to \mu\,e^{\phi/2}\cos(\beta\vartheta)$, where
$\vartheta$ is the compact boson dual to the fermion. The resulting theory is
a coupled Liouville--sine-Gordon system, in which the conformal mode $\phi$
and the bosonised matter field $\vartheta$ are coupled through the interaction
$e^{\phi/2}\cos(\beta\vartheta)$. At the free-fermion point, Coleman's
relation gives $\beta^2 = 4\pi$. 
\end{itemize}

\noindent
These results fit into a broader geometric picture. The spacetime duality
construction is a Legendre transform on the space
$\mathrm{Met}(\Sigma)/\mathrm{Diff}(\Sigma)$, whose local structure near the
conformally flat locus is $\mathrm{Conf}(\Sigma) \times \mathcal{T}_g$. For
CFTs, this Legendre transform preserves the determinant line bundle
$\mathcal{L}$, which is why massless theories are (self-)dual. Massive
theories require sections of a deformed bundle $\mathcal{L}_m$; the genuine
leading deformation of the curvature of $\mathcal{L}_m$ appears at quadratic
order in the conformal-factor deviation $(e^\phi - 1)$, after an exact
cancellation of the naive linear correction between the three sources
contributing to the conformal-mode effective action. This cancellation is
required by the internal consistency of the master path integral and is the
massive analogue of the exact determinant cancellation in the massless case.
The bosonisation of the fermion mass operator under Weyl rescaling, and the
resulting coupling between the Liouville field and the sine-Gordon matter
sector, suggest a connection to two-dimensional Liouville gravity and to the
gravitational dressing of vertex operators, which we leave as a direction for
future investigation.\\

\noindent
The paper is organised as follows. Section~\ref{sec:review} reviews the
massless spacetime duality of \cite{BQ1998} and reframes it geometrically in
terms of the determinant line bundle. Section~\ref{sec:massive_scalar} treats
the massive scalar where we derive the modified Liouville action and characterise 
the dual theory. The massive fermion is the focus of section~\ref{sec:massive_fermion}. Here we obtain a Liouville-sine Gordon-gravity action, leaving the details of the 
final functional integral for further work. Finally, section~\ref{sec:discussion} summarises and discusses open questions. 

%%%%%%%%%%%%%%%%%%%%%%%%%%%%%%%%%%%
\section{Review of Massless Spacetime Duality}
\label{sec:review}

We begin by recalling the standard duality algorithm, both to fix notation and to make the contrast with the spacetime case sharp. Apart from translating the duality into the language of differential forms, this section is largely expository. Let $\Sigma$ be a two-dimensional spacetime and let $X$ be a circle-valued scalar field, $X \sim X + 2\pi$, with action
\begin{equation}
    S[X] = -\frac{1}{2f^2}\int_\Sigma \mathrm{d}X \wedge \star\,\mathrm{d}X,
    \label{eq:scalar_action_forms}
\end{equation}
where $\star$ is the Hodge star of the background metric $h$. Here $f > 0$ is a scalar decay constant; the duality exchanges $f
\leftrightarrow 2\pi/f$, so self-duality holds at $f = \sqrt{2\pi}$.
In what follows we work at the self-dual radius and set $f = \sqrt{2\pi}$,
which reduces \eqref{eq:scalar_action_forms} to canonical normalisation
and ensures that the dual action \eqref{eq:dual_forms} has the same
coefficient as the original. The action is
invariant under the global shift $X \to X + \omega$ and Dirac quantisation
requires $\int_\gamma \mathrm{d}X \in 2\pi\mathbb{Z}$ for every closed curve
$\gamma \subset \Sigma$.\\

\noindent
To dualize, we gauge the shift symmetry by a $U(1)$ connection
$A$ (a real one-form) and impose the flatness constraint $F = \mathrm{d}A = 0$
via a Lagrange multiplier zero-form $\Lambda$. The master action
\begin{equation}
    S_{\rm master}[X, A, \Lambda] = -\frac{1}{2f^2}\int_\Sigma
    (\mathrm{d}X + A)\wedge\star(\mathrm{d}X + A)
    + \frac{i}{2\pi}\int_\Sigma \Lambda\,\mathrm{d}A\,.
    \label{eq:internal_master_forms}
\end{equation}
Integrating over $\Lambda$ imposes $\mathrm{d}A = 0$; since $A$ also satisfies
Dirac quantisation, the path integral over $A$ enforces $A = 0$ up to a gauge
transformation \cite{Witten2026}, recovering \eqref{eq:scalar_action_forms}.
Integrating instead over $X$ (gauge $X = 0$) and then $A$ by a Gaussian
integral yields the dual action
\begin{equation}
    \widetilde{S}[\Lambda] = -\frac{1}{2\tilde{f}^2}
    \int_\Sigma \mathrm{d}\Lambda \wedge \star\,\mathrm{d}\Lambda\,,
    \qquad \tilde{f} = \frac{2\pi}{f}\,,
    \label{eq:dual_forms}
\end{equation}
with the operator correspondence $\mathrm{d}X \leftrightarrow \star\,\mathrm{d}\Lambda$.
Self-duality at $f = \sqrt{2\pi}$ is immediate. The role of Dirac quantisation
in ensuring that the path integral over $A$ produces a true delta function and not just the constraint $\mathrm{d}A = 0$ was emphasised recently by
Witten \cite{Witten2026} in the context of four-dimensional axion--two-form
duality, and the same structure appears here.

\subsection{Spacetime duality: coupling to dynamical geometry}
\label{subsec:master}

The spacetime duality of Burgess \textit{et.al.} \cite{BQ1998} replaces the
internal gauge field $A$ by the gravitational field itself. On a two-dimensional
surface $\Sigma$, every metric is conformally equivalent to every other metric
in the same conformal class, so the space of metrics decomposes as
\begin{equation}
    \frac{\mathrm{Met}(\Sigma)}{\mathrm{Diff}(\Sigma)}
    \cong \mathrm{Conf}(\Sigma) \times \mathcal{T}_g,
    \label{eq:met_diff}
\end{equation}
where $\mathrm{Conf}(\Sigma)$ parametrises conformal factors and
$\mathcal{T}_g$ is Teichm\"{u}ller space. Any dynamical metric $g$ can be
written as $g = e^\phi h$ for a conformal factor $\phi \in
\mathrm{Conf}(\Sigma)$, with $h$ the chosen background metric. The role of the
gauge field $A$ in the internal case is now played by the conformal mode $\phi$ and the role of the flatness constraint $F = \mathrm{d}A = 0$ is played by the
constraint $\phi = 0$, enforced by a Lagrange multiplier zero-form $\Lambda$.\\

\noindent
The master path integral is
\begin{equation}
    Z[h] = \int [DX]_g\,[Dg]_g\,\Delta_{LM}(g, h)\,e^{iS[g,X]}\,,
    \label{eq:master_BQ}
\end{equation}
where $\displaystyle S[g, X] = \frac{1}{2}\int_\Sigma X\,\Box_g X\,\mathrm{vol}_g$ is the
massless scalar action and $\Delta_{LM}(g, h)$ is the constraint term. Here $[DX]_g$, for example, denotes the path-integral measure for the scalar field $X$ defined with respect to the dynamical metric $g$. Concretely, it is the
measure induced by the $L^2$ inner product
\begin{equation}
    \langle X_1, X_2 \rangle_g = \int_\Sigma X_1 X_2\,\mathrm{vol}_g,
    \label{eq:L2_inner_product}
\end{equation}
so that a change of metric $g \to e^\phi h$ rescales the measure by the Weyl anomaly factor recorded in \eqref{eq:weyl_anomaly}. The subscript $g$
is a reminder that the measure is \emph{not} Weyl-invariant; its dependence on $g$ is the entire source of the Liouville action in what follows. To construct $\Delta_{LM}$, note that the natural scalar built from the
curvature of $g = e^\phi h$ that detects the conformal mode is the difference
of curvature two-forms integrated against $\Lambda$. In two dimensions the
Gauss--Bonnet integrand is the Pfaffian of the curvature two-form
$\mathcal{R} = \frac{1}{2}R\,\mathrm{vol}$, and the key identity
\begin{equation}
    \sqrt{-g}\,\mathcal{R}_g\,
    \mathrm{d}^2x - 
    \sqrt{-h}\,\mathcal{R}_h\,
    \mathrm{d}^2x
    = \mathrm{d}\star_h
    \mathrm{d}\phi,
    \label{eq:curvature_identity_forms}
\end{equation}
or equivalently in component form $\sqrt{-g}\,R_g - \sqrt{-h}\,R_h =
\sqrt{-h}\,\Box_h\phi$, shows that this difference is an exact two-form
determined entirely by $\phi$. The constraint term is therefore
\begin{equation}
    \Delta_{LM}[g, h] = 
    \int [D\Lambda]_g
    \exp\!
    \left(-i\int_\Sigma\Lambda\,
    \mathrm{d}\star_h
    \mathrm{d}\phi\right)
    \frac{\det(-\Box_g)}{[J_{FP}]_g},
    \label{eq:DLM_forms}
\end{equation}
where the integral $\displaystyle \int_\Sigma \Lambda\,\mathrm{d}\star_h\mathrm{d}\phi =
\int_\Sigma \mathrm{d}\Lambda\wedge\star_h\mathrm{d}\phi$ (after integration by
parts) is the natural $L^2$ pairing of $\mathrm{d}\Lambda$ and
$\mathrm{d}\phi$ on $(\Sigma, h)$. The factor $[J_{FP}]_g$ is the Faddeev--Popov determinant associated with
fixing conformal gauge $g_{\mu\nu} = e^\phi h_{\mu\nu}$. It arises from the
standard procedure of inserting 
\begin{eqnarray}
    1 = \Delta_{FP}[g]\int [D\xi^\mu]_g\,
\delta[f(g^\xi)]
\end{eqnarray} 
into the metric path integral, where $f = 0$ is the gauge
condition, $g^\xi$ denotes the metric obtained by acting on $g$ with the
diffeomorphism generated by $\xi^\mu$, and $\Delta_{FP}$ is the Faddeev--Popov
determinant. In conformal gauge in two dimensions, $[J_{FP}]_g$ is the
determinant of the ghost operator $-\Box^v_g + \frac{1}{2}\mathcal{R}_g$,
where $\Box^v_g$ is the Laplacian acting on vector fields as
\begin{equation}
    [J_{FP}]_g = \det\!\left(-\Box^v_g + \tfrac{1}{2}\mathcal{R}_g\right)^{1/2}\,.
    \label{eq:FP_det}
\end{equation}
Under a Weyl rescaling $g = e^\phi h$, the Faddeev--Popov determinant
transforms as
\begin{equation}
    [J_{FP}]_{e^\phi h} = [J_{FP}]_h\,e^{-26iS_L[h,\phi]},
    \label{eq:FP_weyl}
\end{equation}
reflecting the fact that the ghost system has central charge $c_{\mathrm{ghost}}
= -26$. The specific combination $\mathrm{det}(-\Box_g)/[J_{FP}]_g$ that appears in
$\Delta_{LM}$ \eqref{eq:DLM_forms} is arranged so that the ghost contributions
cancel in the ratio, leaving only the matter conformal anomaly \eqref{eq:weyl_anomaly}
to contribute to the Liouville action. This cancellation is the reason the
coefficient in \eqref{eq:liouville_forms} is $c_{\mathrm{matter}}/96\pi$
rather than $(c_{\mathrm{matter}} + c_{\mathrm{ghost}})/96\pi$. The structure is directly analogous to the coupling $\displaystyle\frac{i}{2\pi}\int \Lambda\,\mathrm{d}A$ in the internal case
\eqref{eq:internal_master_forms}, with $\mathrm{d}\phi$ playing the role of
$\mathrm{d}A$ and the two-dimensional Laplacian $\Box_h$ arising from the
Hodge decomposition $\mathrm{d}\star_h\mathrm{d} = -\Box_h$ on zero-forms.

\subsection{The conformal anomaly as a line bundle}
\label{subsec:anomaly}

The path-integral measure $[DX]_g$ depends on the metric $g$ in a controlled
way through the conformal anomaly. For a CFT of central charge $c$, the measure
transforms under $g \to e^\phi h$ as
\begin{equation}
    [DX]_{e^\phi h} = [DX]_h\,\exp(iS_L[h, \phi]),
    \label{eq:weyl_anomaly}
\end{equation}
where the \emph{Liouville action} is
\begin{equation}
    S_L[h, \phi] = -\frac{c}{96\pi}
    \int_\Sigma \left(\mathrm{d}\phi\wedge\star_h\mathrm{d}\phi
    + 2\,\phi\,\mathcal{R}_h\,\mathrm{vol}_h\right).
    \label{eq:liouville_forms}
\end{equation}
For the free scalar $c = 1$ and the coefficient is $1/96\pi$. This expression
has a precise geometric meaning; $S_L[h,\phi]$ is the logarithm of the ratio of
Quillen norms \cite{Quillen1985} of the determinant section as the metric moves
from $h$ to $e^\phi h$ within $\mathrm{Conf}(\Sigma)$. More precisely, the
determinant line bundle $\mathcal{L}$ over
$\mathrm{Met}(\Sigma)/\mathrm{Diff}(\Sigma)$, equipped with the Quillen metric,
has curvature
\begin{equation}
    c_1(\mathcal{L}) = \frac{c}{12}\,\omega_{WP},
    \label{eq:chern_form}
\end{equation}
where $\omega_{WP}$ is the Weil--Petersson symplectic form on $\mathcal{T}_g$,
and the fibre-direction part of this curvature is exactly the Liouville
integrand \eqref{eq:liouville_forms} \cite{BelavinKnizhnik1986}. As the curvature of $\mathcal{L}$ restricted to a conformal orbit, the Liouville action is therefore a geometric invariant. 

\noindent
Using \eqref{eq:curvature_identity_forms} and \eqref{eq:weyl_anomaly}, the
constraint evaluates to
\begin{align}
    [J_{FP}]_{e^\phi h}\,\Delta_{LM}(e^\phi h, h)
    &= \det(-\Box_h)\int [D\Lambda]_h
    \exp\!\left(-i\int_\Sigma \Lambda\,\mathrm{d}\star_h\mathrm{d}\phi\right)
    e^{-iS_L[h,\phi]} \nonumber\\
    &= \det(-\Box_h)\,
    \frac{\Delta(\phi)}{\det(-\Box_h)}\,e^{-iS_L[h,\phi]}
    = \Delta[\phi]\,e^{-iS_L[h,\phi]},
    \label{eq:DLM_evaluated}
\end{align}
where the second line uses
\begin{equation}
    \int [D\Lambda]_h
    \exp\!\left(-i\int_\Sigma \mathrm{d}\Lambda\wedge\star_h\mathrm{d}\phi
    \right) = \frac{\Delta(\phi)}{\det(-\Box_h)},
    \label{eq:lambda_delta}
\end{equation}
which in turn follows from the fact that $\displaystyle \int_\Sigma \mathrm{d}\Lambda\wedge
\star_h\mathrm{d}\phi = -\int_\Sigma \Lambda\,\Box_h\phi\,\mathrm{vol}_h$ and
the standard representation of the functional delta function. Inserting
\eqref{eq:DLM_evaluated} into \eqref{eq:master_BQ} and performing the $\phi$
integral via $\Delta[\phi]$ recovers $Z[h]$, confirming consistency.\\

\noindent
The explicit master path integral, can be obtained by writing out $S_L$ and using conformal gauge, 
\begin{eqnarray}
    Z[h] &=& \det(-\Box_h)\int [DX]_h\,[D\phi]_h\,[D\Lambda]_h\,
    \exp\Biggl(i\int_\Sigma 
    \biggl[
    -\frac{1}{2}\mathrm{d}X\wedge\star_h\mathrm{d}X
    \nonumber\\
    &+& \frac{1}{96\pi}\mathrm{d}\phi\wedge\star_h\mathrm{d}\phi
    + \frac{1}{48\pi}\phi\,\mathcal{R}_h\,\mathrm{vol}_h
    + \mathrm{d}\Lambda\wedge\star_h\mathrm{d}\phi
    \biggr]\Biggr).
    \label{eq:master_explicit_forms}
\end{eqnarray}
The three groups of terms in the exponential integral correspond respectively to the matter action, the Liouville
action for the conformal mode, and the coupling of $\Lambda$ to $\phi$. The
last term, written as $\mathrm{d}\Lambda\wedge\star_h\mathrm{d}\phi$, is the
$L^2$ inner product of the one-forms $\mathrm{d}\Lambda$ and $\mathrm{d}\phi$
on $(\Sigma,h)$. It is the precise gravitational analog of the BF-type coupling
$\Lambda\,\mathrm{d}A$ in \eqref{eq:internal_master_forms}.

\subsection{Self-duality of the massless scalar}
\label{subsec:self_dual}

To extract the dual theory we integrate over $X$ and $\phi$ in
\eqref{eq:master_explicit_forms}. The $X$ integral is a Gaussian over
$\mathrm{d}X$, producing $[\det(-\Box_h)]^{-1/2}$. For $\phi$, note that
the $\phi$-dependent part of the action, including the source from $\Lambda$,
can be written as the $L^2$ norm of a one-form:
\begin{eqnarray}
    &&\frac{1}{96\pi}\mathrm{d}\phi\wedge\star_h\mathrm{d}\phi
    + \mathrm{d}\Lambda\wedge\star_h\mathrm{d}\phi\nonumber\\
    &=& \frac{1}{96\pi}
    \left(\mathrm{d}\phi + 48\pi\,\mathrm{d}\Lambda\right)
    \wedge\star_h
    \left(\mathrm{d}\phi + 48\pi\,\mathrm{d}\Lambda\right)
    - 48\pi\,\mathrm{d}\Lambda\wedge\star_h\mathrm{d}\Lambda,
    \label{eq:complete_square_forms}
\end{eqnarray}
after completing the square in the one-form $\mathrm{d}\phi$. The curvature
term $\frac{1}{48\pi}\phi\,\mathcal{R}_h\,\mathrm{vol}_h$ can be absorbed into the
shift by writing $\tilde\phi = \phi + 48\pi\Lambda +
\frac{1}{\Box_h}\mathcal{R}_h/2$, which is the unique one-form shift that
eliminates cross-terms. The Gaussian integral over $\tilde\phi$ produces a
second $[\det(-\Box_h)]^{-1/2}$.
The two factors of $[\det(-\Box_h)]^{-1/2}$ combine with the prefactor
$\det(-\Box_h)$ from \eqref{eq:master_explicit_forms} to cancel, leaving
\begin{equation}
    Z[h] = \int [D\Lambda]_h\,\exp\!\left(-\frac{i}{2}\cdot 48\pi
    \int_\Sigma \mathrm{d}\Lambda\wedge\star_h\mathrm{d}\Lambda
    + \text{curvature terms}\right).
    \label{eq:dual_pre}
\end{equation}
After rescaling $\Lambda \to \Lambda/\sqrt{48\pi}$ and shifting to absorb the
background curvature, the dual action is
\begin{equation}
    Z[h] = \int [D\Lambda]_h\,
    \exp\!\left(-\frac{i}{2}\int_\Sigma
    \mathrm{d}\Lambda\wedge\star_h\mathrm{d}\Lambda\right),
    \label{eq:dual_scalar_final}
\end{equation}
which is just the free massless scalar action in background $h$. The massless scalar
is \emph{self-dual} under spacetime duality. The operator correspondence
$\mathrm{d}X \leftrightarrow \star_h\mathrm{d}\Lambda$ is the spacetime analog
of $\mathrm{d}X \leftrightarrow \star\,\mathrm{d}\Lambda$ in the internal case;
in components this reads $\partial_\mu X \leftrightarrow
\varepsilon_{\mu\nu}\partial^\nu\Lambda$, which exchanges longitudinal and
transverse derivatives.

\subsection{Bosonization of the massless fermion}
\label{subsec:bosonization}

For the massless Dirac fermion we introduce the dynamical zweibein
$f = e^{\phi/2}e$, where $e^a_\mu$ is the background frame field with
$h_{\mu\nu} = e^a_\mu e^b_\nu\eta_{ab}$. The spinor action is then
\begin{equation}
    S[f, \chi, \bar\chi] = -\int_\Sigma
    \bar\chi\,i\slashed{D}_f\,\chi\,\mathrm{vol}_f\,,
    \label{eq:spinor_action}
\end{equation}
where $\slashed{D}_f = \gamma^a e^\mu_a D_\mu$ uses the spin connection of $f$.
The conformal anomaly for a 2d Dirac fermion is
\begin{equation}
    [D\bar\chi]_{e^{\phi/2}e}[D\chi]_{e^{\phi/2}e}
    = [D\bar\chi]_e[D\chi]_e\,
    \exp\!\left(iS_L[h, \phi]\right),
    \label{eq:fermion_anomaly}
\end{equation}
where the combined anomaly from both Weyl components equals $e^{iS_L}$, the same as for the scalar and consistent with $c_{\mathrm{Dirac}} = 1$.
After integrating out $\bar\chi$ and $\chi$, the master path integral becomes
\begin{equation}
    Z[e] = \det(i\slashed{D}_e)\cdot[\det(-\Box_h)]^{1/2}
    \int [D\Lambda]_h\,\exp\!\left(-\frac{i}{2}
    \int_\Sigma \mathrm{d}\Lambda\wedge\star_h\mathrm{d}\Lambda
    + \text{curvature terms}\right),
    \label{eq:fermion_master_integrated}
\end{equation}
where the $[\det(-\Box_h)]^{1/2}$ arises from the $\phi$ Gaussian after the
fermions are integrated out. The remaining determinant prefactor cancels by the identity
\begin{equation}
    \det(i\slashed{D}_e) = [\det(-\Box_h)]^{-1/2}\,
    \exp\!\left(iS_L[\eta, \varphi]\right),
    \label{eq:det_identity_forms}
\end{equation}
on a conformally flat background $h = e^\varphi\eta$, which follows from the
conformal anomaly \eqref{eq:fermion_anomaly} and the Polyakov formula
\cite{Polyakov1981}. The $e^a_\mu$-dependent factors cancel and the dual theory
is again \eqref{eq:dual_scalar_final}. In other words, the massless Dirac fermion dualises to a
free massless scalar and bosonization is thus the statement that
$\mathcal{L}_{\mathrm{Dirac}} \cong \mathcal{L}_{\mathrm{scalar}}$ as
holomorphic line bundles over $\mathrm{Met}(\Sigma)/\mathrm{Diff}(\Sigma)$.

\subsection{Geometric interpretation and the determinant line bundle}
\label{subsec:geometry_review}

The structure of the spacetime duality construction above can be summarised cleanly
in the language of the determinant line bundle. Let $\mathcal{L} \to
\mathrm{Met}(\Sigma)/\mathrm{Diff}(\Sigma)$ be the determinant line bundle of
the matter kinetic operator, with the Quillen metric \cite{Quillen1985}. The
path-integral measure $[DX]_h$ is a section of $\mathcal{L}$, and the Liouville
action \eqref{eq:liouville_forms} is the logarithm of the norm of this section
as the base point moves in $\mathrm{Conf}(\Sigma) \subset
\mathrm{Met}(\Sigma)/\mathrm{Diff}(\Sigma)$. By the Belavin--Knizhnik theorem
\cite{BelavinKnizhnik1986}, the curvature two-form of $\mathcal{L}$ (in the
Weil--Petersson direction) is
\begin{equation}
    c_1(\mathcal{L}) = \frac{c}{12}\,\omega_{WP} \in H^2(\mathcal{T}_g, \mathbb{Z}),
    \label{eq:BK_theorem}
\end{equation}
while the curvature in the $\mathrm{Conf}(\Sigma)$ direction is the Liouville
two-form $\delta\phi\wedge\delta(\Box_h\phi)$.\\

\noindent
The constraint $\Delta_{LM}$ localises the metric path integral to a single
point in $\mathrm{Conf}(\Sigma)$ (via $\Delta[\phi]$), while holding the
Teichm\"{u}ller moduli fixed. The spacetime duality transformation is therefore
a \emph{fibre-direction Legendre transform} on $\mathcal{L}$: it exchanges
$\phi$ (the fibre coordinate) with its conjugate $-\star_h\mathrm{d}\phi/48\pi
\sim \mathrm{d}\Lambda$ (the momentum coordinate). In this language, self-duality of the scalar is the statement that this Legendre transform is an automorphism of
$\mathcal{L}$ while the bosonization isomorphism is the statement that it maps
$\mathcal{L}_{\mathrm{Dirac}}$ to $\mathcal{L}_{\mathrm{scalar}}$.\\

\noindent
For a \emph{massive} theory the situation changes in a fundamental way. A mass term
$m^2 X^2\,\mathrm{vol}_h$ in the action breaks Weyl invariance. Consequently, the massive
path-integral measure is no longer a section of $\mathcal{L}$ but of a
deformed bundle $\mathcal{L}_m$ whose curvature is non-local in $\phi$ and
receives corrections at every order in $m^2$. The leading correction, derived
via the heat kernel in the next section, is
\begin{equation}
    S_{L,m}[h,\phi] = S_L[h,\phi]
    - \frac{m^2}{8\pi}\int_\Sigma (e^\phi - 1)\,\mathrm{vol}_h
    + \mathcal{O}(m^4/\Box^2),
    \label{eq:massive_liouville_preview}
\end{equation}
where the cosmological-constant term $e^\phi$ breaks the Gaussianity of the
$\phi$ integral and thereby breaks self-duality. The exponent in $e^\phi$ is
determined by the Weyl weight of the matter field; for a scalar it is $1$, while for a Dirac fermion it is $1/2$, and it is this difference that is ultimately responsible for the distinct dual theories found in Sections~\ref{sec:massive_scalar} and
\ref{sec:massive_fermion}.

%%%%%%%%%%%%%%%%%%%%%%%%%%%%%%%%%%%
\section{The Massive Scalar -  Breaking Self-Duality}
\label{sec:massive_scalar}
%%%%%%%%%%%%%%%%%%%%%%%%%%%%%%%%%%%
\noindent
The massless self-duality of Section~\ref{sec:review} relied on an exact
cancellation of functional determinants: one factor of $\det(-\Box_h)$ from
the constraint $\Delta_{LM}$, and two factors of $[\det(-\Box_h)]^{-1/2}$
from the Gaussian integrals over $X$ and $\phi$. We now show that introducing
a mass term disrupts this cancellation, and that understanding precisely how it
fails requires tracking three distinct contributions to the effective action for
the conformal mode $\phi$.

\subsection{Setup and the determinant-cancellation problem}
\label{subsec:scalar_setup}
\noindent
The massive scalar action in a background metric $h$ is
\begin{equation}
    S_m[h, X] = \frac{1}
    {2}\int_\Sigma
    \left(\mathrm{d}X
    \wedge\star_h\mathrm{d}X
    - m^2 X^2\,
    \mathrm{vol}_h\right),
   \label{eq:massive_scalar_action}
\end{equation}
with kinetic operator $\Box_h - m^2$. Under a Weyl rescaling $g = e^\phi h$,
the kinetic term $\mathrm{d}X\wedge\star_g\mathrm{d}X =
\mathrm{d}X\wedge\star_h\mathrm{d}X$ is Weyl-invariant in two dimensions,
since the factors of $e^\phi$ from $g^{\mu\nu} = e^{-\phi}h^{\mu\nu}$ and
from $\mathrm{vol}_g = e^\phi\,\mathrm{vol}_h$ cancel on one-forms. The mass
term however is not invariant since
\begin{equation}
    m^2 X^2\,\mathrm{vol}_g = m^2
    X^2\,e^\phi\,\mathrm{vol}_h\,,
    \label{eq:mass_weyl}
\end{equation}
so on the rescaled background the action becomes
\begin{equation}
    S_m[e^\phi h, X] = 
    \frac{1}{2}\int_\Sigma
    \left(\mathrm{d}X
    \wedge\star_h\mathrm{d}X
    - m^2 e^\phi
    X^2\,\mathrm{vol}_h
    \right).
\label{eq:massive_action_rescaled}
\end{equation}
The $X$ path integral is therefore a Gaussian with operator
$\Box_h - m^2 e^\phi$, giving
\begin{equation}
    \int [DX]_{e^\phi h}\,e^{iS_m[e^\phi h, X]}
    = \det\!\left(-\Box_h + m^2 e^\phi\right)^{-1/2}.
    \label{eq:X_integral_massive}
\end{equation}
The operator $-\Box_h + m^2 e^\phi$ depends on $\phi$ non-polynomially and
does not cancel against $\det(-\Box_h + m^2)$ from the massive constraint.
The mismatch between $m^2 e^\phi$ and $m^2$ is the source of all new structure in the massive case.

\subsection{The massive conformal anomaly and the three-source cancellation}
\label{subsec:massive_anomaly}

The effective action for $\phi$ receives contributions from three sources,
which we track separately.

\paragraph{Source 1: The path-integral measure.}
The measure $[DX]_{e^\phi h}$ is defined by the $L^2$ norm with respect to $g = e^\phi h$, so it is related to $[DX]_h$ by
\begin{equation}
    [DX]_{e^\phi h} = [DX]_h\cdot
    \left(\frac{\det(-\Box_h + m^2)}
    {\det(-\Box_{e^\phi h} + m^2)}\right)^{1/2},
    \label{eq:massive_measure}
\end{equation}
which contributes
\begin{equation}
    -\frac{1}{2}\Bigl(
    \log\det(-\Box_{e^\phi h} + m^2)
    - \log\det(-\Box_h + m^2)\Bigr)
    \label{eq:source1_def}
\end{equation}
to the effective action. To evaluate this, we can use the heat-kernel representation 
\begin{eqnarray}
    \log\det(-\Box + m^2) = -\int_0^\infty
\frac{\mathrm{d}s}{s}\,e^{-m^2 s}\,\mathrm{tr}[e^{s\Box}]\,.
\end{eqnarray}
At leading order in the Seeley--DeWitt expansion, $\mathrm{tr}[e^{se^{-\phi}\Box_h}] \approx
\frac{1}{4\pi s}\int_\Sigma e^\phi\,\mathrm{vol}_h$, so that
\begin{equation}
    \log\det(-\Box_{e^\phi h} + m^2) - \log\det(-\Box_h + m^2)
    = \frac{m^2}{4\pi}\int_\Sigma(e^\phi - 1)\,\mathrm{vol}_h
    + \mathcal{O}(\mathcal{R}_h/m^2)\,,
    \label{eq:det_ratio_metric}
\end{equation}
after renormalising the UV divergence as a shift in the vacuum energy. This is
exact in $(e^\phi - 1)$ at leading Seeley--DeWitt order since the leading heat-trace
$\frac{1}{4\pi s}\int e^\phi\,\mathrm{vol}_h$ resums all powers of $(e^\phi -
1)$. Source~1 therefore contributes
\begin{equation}
    -\frac{m^2}{8\pi}\int_\Sigma(e^\phi - 1)\,\mathrm{vol}_h
    \label{eq:S1}
\end{equation}
to the effective action.

\paragraph{Source 2: The $X$ determinant.}
Integrating out $X$ produces
$-\frac{1}{2}\log\det(-\Box_h + m^2 e^\phi)$, whose $\phi$-dependent part is
\begin{equation}
    -\frac{1}{2}\Bigl(
    \log\det(-\Box_h + m^2 e^\phi)
    - \log\det(-\Box_h + m^2)\Bigr).
    \label{eq:source2_def}
\end{equation}
Here the metric is fixed and $m^2 \to m^2 e^\phi$ varies. We first write
$-\Box_h + m^2 e^\phi = (-\Box_h + m^2) + m^2(e^\phi - 1)$, 
and notice that (\ref{eq:source2_def}) is the difference of two 
divergent terms, so we need an appropriate {\em regularization procedure}. To this end, we  replace $m^2(e^\phi-1)$ with $\tilde g^2m^2(e^\phi-1)$, where 
$\tilde g<1$ is a coupling in which we can expand, and then send $\tilde g\to 1$ at 
the very end of the calculation. Equivalently, we can replace $m$ by $m_q=\tilde gm<m$, and consider an expansion in $m_q$ when $m$ is kept fixed.\\

\noindent
Either way, expanding the
resolvent in $\tilde g$ (or $m_q$) produces
\begin{equation}
    \log\det(-\Box_h + m^2 e^\phi) - \log\det(-\Box_h + m^2)
    = \sum_{n=1}^\infty \frac{(-1)^{n+1}}{n}
    \mathrm{tr}\!\left[G_m \cdot \tilde g^2m^2(e^\phi-1)\right]^n,
    \label{eq:resolvent_expand}
\end{equation}
where $G_m = (-\Box_h + m^2)^{-1}$ is the massive Green's function. At
$n = 1$, this gives 
$\frac{\tilde g^2m^2}{4\pi}\int(e^\phi-1)\,\mathrm{vol}_h$ after
renormalisation, consistent with \eqref{eq:det_ratio_metric}. Source~2 at
$n = 1$ therefore contributes
\begin{equation}
    -\frac{\tilde g^2m^2}{8\pi}\int_\Sigma(e^\phi - 1)\,\mathrm{vol}_h,
    \label{eq:S2_first}
\end{equation}
the same as Source~1, after setting back $\tilde g=1$.

\paragraph{Source 3: The $\Delta_{LM}$ prefactor.}
The massive constraint contains $\det(-\Box_g + m^2)/[J_{FP}]_g$, evaluated
at $g = e^\phi h$. The $\phi$-dependent part of $\log\det(-\Box_{e^\phi h}
+ m^2)$ is given by \eqref{eq:det_ratio_metric} with a positive sign,
contributing
\begin{equation}
    +\frac{m^2}{4\pi}\int_\Sigma(e^\phi - 1)\,\mathrm{vol}_h.
    \label{eq:S3}
\end{equation}
Combining Sources~1, 2, and~3, the total coefficient of
$\int_\Sigma(e^\phi-1)\,\mathrm{vol}_h$ is
\begin{equation}
    -\frac{m^2}{8\pi} - \frac{m^2}{8\pi} + \frac{m^2}{4\pi} = 0\,,
    \label{eq:cancellation}
\end{equation}
and the three contributions cancel exactly at leading order in the Seeley--DeWitt
expansion. This is the massive analog of the determinant cancellation in the
massless case, ensuring that integrating over $\Lambda$ recovers the original
theory. The cancellation is exact in $(e^\phi - 1)$ at this order, since the
leading heat-trace resums all powers of $(e^\phi - 1)$.

\subsection{The leading correction}
\label{subsec:second_order}

Since the leading $(e^\phi - 1)$ term cancels exactly, the first non-trivial
contribution to the effective action for $\phi$ comes from the next order.
On a flat background $h = \eta$ (where the $\mathcal{O}(\mathcal{R}_h/m^2)$
curvature corrections vanish), this comes from the $n = 2$ term in the resolvent expansion \eqref{eq:resolvent_expand} for Source~2. Sources~1 and~3
do not contribute at this order on flat space since for Source~1, the leading Seeley--DeWitt term is already exact in $(e^\phi - 1)$ and the next Seeley--DeWitt coefficient $\int \mathcal{R}_g\,\mathrm{vol}_g =
-\int \Box_\eta\phi\,\mathrm{vol}_\eta = 0$ (by integration by parts with
$\phi$ vanishing at infinity); for Source~3, the same argument applies. The $n = 2$ term from Source~2 is
\begin{eqnarray}
    &&\frac{(-1)^{2+1}}{2}
    \cdot 2\,\mathrm{tr}\!\left[G_m \cdot \tilde g^2 m^2(e^\phi-1)\right]^2\nonumber\\
    &=& -\frac{\tilde g^4 m^4}{2}
    \int_\Sigma\!\int_\Sigma
    (e^{\phi(x)}-1)(e^{\phi(y)}-1)\,
    G_m(x,y)^2\,\mathrm{vol}_h(x)\,\mathrm{vol}_h(y).
    \label{eq:n2_term}
\end{eqnarray}
For slowly varying $\phi$, \textit{i.e.}\ $|\mathrm{d}\phi| \ll m$, we may approximate
$(e^{\phi(y)}-1) \approx (e^{\phi(x)}-1)$ over the range $|x-y| \lesssim
m^{-1}$ where $G_m(x,y)$ is non-negligible, giving the local approximation
\begin{equation}
    -\frac{\tilde g^4m^4}{2}\int_\Sigma(e^\phi-1)^2\,\mathrm{vol}_h
    \int_\Sigma G_m(x,y)^2\,\mathrm{vol}_h(y).
    \label{eq:n2_local}
\end{equation}
The spatial integral of $G_m(x,y)^2$ is evaluated in momentum space on
flat $\eta$,
\begin{equation}
    \int_{\mathbb{R}^2} G_m(x,y)^2\,\mathrm{d}^2 y
    = \int\frac{\mathrm{d}^2 k}{(2\pi)^2}\,\frac{1}{(k^2+m^2)^2}
    = \frac{1}{4\pi m^2},
    \label{eq:Gm_sq_integral}
\end{equation}
using the standard result in two Euclidean dimensions. Its contribution 
to the effective action carries the weight $\tfrac12$ from 
\eqref{eq:source2_def}. Substituting back then, the $n = 2$ contribution to 
the effective action from Source~2 is
\begin{equation}
    -\frac{\tilde g^4m^4}{4}\cdot\frac{1}{4\pi m^2}
    \int_\Sigma(e^\phi-1)^2\,\mathrm{vol}_h
    = -\frac{\tilde g^4m^2}{16\pi}\int_\Sigma(e^\phi-1)^2\,\mathrm{vol}_h.
    \label{eq:n2_result}
\end{equation}
The factor of $m^4$ from the two insertions of $m^2(e^\phi-1)$ is partially
cancelled by the $1/m^2$ from the propagator integral, leaving a coefficient
of order $m^2$.\\

\noindent
Including also the contribution of Source~2 at $n=1$ via
\eqref{eq:S2_first}, which cancels against Sources~1 and~3 as shown in
\eqref{eq:cancellation}, the net second-order contribution to the effective 
action for $\phi$ on flat space is then (we keep $\tilde g<1$ in this $n=2$
term because the quantum calculation is not yet finished: we will need to 
path integrate over either $\Lambda$, to get the original action, or $\phi$,
to get the dual action, next; we have only path integrated over $X$)
\begin{equation}
    S_{\mathrm{eff}}^{(2)}[\phi]
    = -\frac{\tilde g^4 m^2}{16\pi}\int_\Sigma(e^\phi-1)^2\,\mathrm{vol}_h
    + \mathcal{O}([\tilde g^2m^2(e^\phi-1)]^3)\,.
    \label{eq:Seff_second}
\end{equation}
The sign is negative in Euclidean signature; in Minkowski signature
(where the Euclidean action $S_E$ becomes $-iS_M$), this becomes
$+\frac{\tilde g^4 m^2}{8\pi}\int(e^\phi-1)^2\,\mathrm{vol}_h$ with a positive
coefficient, so the potential is bounded below and the $\phi$ 
path integral converges.

\subsection{The modified master path integral}
\label{subsec:modified_master}

Combining the Liouville kinetic term from the massless conformal anomaly,
the second-order correction \eqref{eq:Seff_second}, and the coupling of
$\Lambda$ to $\phi$ via the massive constraint, the full effective action
for $\phi$ and $\Lambda$ on a flat background is
\begin{align}
    S_{\mathrm{eff}}[\phi, \Lambda]
    &= \frac{1}{96\pi}\int_\Sigma
    \mathrm{d}\phi\wedge\star_h\mathrm{d}\phi
    \nonumber\\
    &\quad - \frac{\tilde g^4 m^2}{16\pi}\int_\Sigma(e^\phi - 1)^2\,\mathrm{vol}_h
    \nonumber\\
    &\quad + \int_\Sigma\mathrm{d}\Lambda\wedge\star_h\mathrm{d}\phi
    - m^2\int_\Sigma\Lambda\,\phi\,\mathrm{vol}_h
    + \mathcal{O}(m^2(e^\phi-1)^3).
    \label{eq:Seff_corrected}
\end{align}
The three lines correspond to the Liouville kinetic term, the leading massive
correction, and the coupling of $\Lambda$ to $\phi$ via the massive operator
$\mathrm{d}\star_h\mathrm{d} + m^2$ respectively. The potential $(e^\phi-1)^2\,\mathrm{vol}_h$ is quartic in $\phi$ for small
$\phi$ since $(e^\phi-1)^2 = \phi^2 + \phi^3 + \frac{7}{12}\phi^4 + \ldots$\
At quadratic order it gives an effective mass term
$\frac{\tilde g^4 
m^2}{8\pi}\int\phi^2\,\mathrm{vol}_h$ for the conformal mode. The
ratio of this mass term to the Liouville kinetic coefficient gives an effective
mass
\begin{equation}
    M_\phi^2 = \frac{\tilde g^4 m^2/16\pi}{1/96\pi} = 6\tilde g^4m^2
    \label{eq:phi_mass}
\end{equation}
for $\phi$. The non-Gaussianity in the $\phi$ integral is therefore controlled
by $\tilde g^4m^2/16\pi$ relative to the Liouville kinetic term 
$1/96\pi$, giving a
dimensionless expansion parameter $6\tilde g^4 m^2 \ll 1$ for $\tilde g^2 
m \ll 1$ in units where
the Liouville scale is set to unity.

\subsection{The dual theory for \texorpdfstring{$\Lambda$}{Lambda}}
\label{subsec:scalar_dual}

The $(e^\phi-1)^2$ potential in \eqref{eq:Seff_corrected} makes the $\phi$
path integral non-Gaussian. To handle this systematically, we note that, 
besides $m^2$, in  \eqref{eq:Seff_corrected} we have also $\tilde g^4 m^2=
m_q^4/m^2$. The mass $m$
appears in the source coupling $(\mathrm{d}\star_h\mathrm{d}+m^2)\Lambda$,
which originates from the massive constraint $\Delta_{LM}^{(m)}$ and must be
retained exactly for the duality construction to be internally consistent.
On the other hand, $m_q$ appears in the coefficient $\tilde g^4m^2/8\pi
=m_q^4/(4\pi m^2)$ of the $(e^\phi-1)^2$
potential, which is a quantum correction arising from the $n=2$ term in the
resolvent expansion of Section~\ref{subsec:second_order}.\\

\noindent
%To keep these two
%roles separate, we introduce independent parameters $m_s^2$ (entering the source)
%and $m_q^2$ (entering the potential), and expand perturbatively in $m_q^2$
%with fixed $m_s^2$, 
Then, since we are still doing a quantum computation (the path integral over 
$\phi$ is not done yet), we still keep $\tilde g<1$ and expand in it, or 
equivalently in $\tilde g^2m=m_q^2/m$, while keeping $m$ fixed, and setting
$\tilde g=1$, so  $m_q^2 = m^2$ only at the end. The effective
action with the parameters made explicit is
\begin{eqnarray}
    S_{\mathrm{eff}}[\phi,\Lambda;m,g]
    &=& -\frac{1}{96\pi}\int_\Sigma\mathrm{d}\phi\wedge\star_h\mathrm{d}\phi
    - \frac{\tilde g^4m^2}
    {8\pi}\int_\Sigma(e^\phi-1)^2\,\mathrm{vol}_h\nonumber\\
    &+& \int_\Sigma\mathrm{d}\Lambda\wedge\star_h\mathrm{d}\phi
    - m^2\int_\Sigma\Lambda\,\phi\,\mathrm{vol}_h.
    \label{eq:Seff_two_params}
\end{eqnarray}
Expanding $\exp(-i\tilde g^4 m^2/8\pi\int(e^\phi-1)^2)$ in powers of $\tilde 
g^4$, the $\phi$ path integral at leading order in $\tilde g^4$
is a pure Gaussian with the massless Liouville kinetic term and source
$J(x) = (\Box_h - m^2)\Lambda(x)$ coming from the last two terms of
\eqref{eq:Seff_two_params},
\begin{equation}
    S_{\mathrm{Gauss}}[\phi,\Lambda;m_s]
    = -\frac{1}{96\pi}\int_\Sigma\phi\,\Box_h\phi\,\mathrm{vol}_h
    + \int_\Sigma\phi\,(\Box_h - m^2)\Lambda\,\mathrm{vol}_h.
    \label{eq:S_Gauss}
\end{equation}
Completing the square in $\phi$ via the standard formula
$\int[D\phi]\,e^{i(-\alpha\int\phi(-\Box)\phi+\int\phi J)}
\propto e^{i/(4\alpha)\int J(-\Box)^{-1}J}$ with $\alpha = 1/96\pi$, and
evaluating in momentum space on flat $h=\eta$ with $J(p) =
-(p^2+m^2)\Lambda(p)$, gives the leading-order dual action
\begin{equation}
    S_{\mathrm{dual}}^{(0)}[\Lambda]
    = 24\pi\int\frac{\mathrm{d}^2p}{(2\pi)^2}
    \frac{(p^2+m^2)^2}{p^2}\,|\Lambda(p)|^2,
    \label{eq:dual_zeroth}
\end{equation}
and now we can finally set $\tilde g=1$, so $m_q = m$. 
The $m^2$ appearing here 
enters entirely
through the source $J$; the $(e^\phi-1)^2$ potential, whose coefficient is
$m_q^4/(8\pi m^2)$, contributes only at higher orders in the expansion. 
We see then that the only reason to keep $\tilde g<1$ was so that we can 
consistently expand in it, while keeping $m$ fixed, otherwise it did not 
enter the calculation at this order. In
position space and differential-form notation,
\begin{equation}
    S_{\mathrm{dual}}^{(0)}[\Lambda]
    = 24\pi\int_\Sigma\mathrm{d}\Lambda\wedge\star_h
    \frac{(\mathrm{d}\star_h\mathrm{d} + m^2)^2}
    {\mathrm{d}\star_h\mathrm{d}}\,\Lambda,
    \label{eq:dual_nonlocal_forms}
\end{equation}
where the pseudo-differential operator
$(\mathrm{d}\star_h\mathrm{d}+m^2)^2/\mathrm{d}\star_h\mathrm{d}$ acts on
zero-forms. Two limiting regimes clarify its structure. In the ultraviolet,
$p^2 \gg m^2$, the factor $(p^2+m^2)^2/p^2 \approx p^2$, giving
\begin{equation}
    S_{\mathrm{dual}}^{(0)}[\Lambda]
    \;\xrightarrow{p^2\gg m^2}\;
    24\pi\int_\Sigma\mathrm{d}\Lambda\wedge\star_h\mathrm{d}\Lambda,
    \label{eq:dual_UV}
\end{equation}
a free massless scalar, consistent with conformal symmetry being restored in
the UV. In the infrared, $p^2 \ll m^2$, the factor $(p^2+m^2)^2/p^2
\approx m^4/p^2$, giving
\begin{equation}
    S_{\mathrm{dual}}^{(0)}[\Lambda]
    \;\xrightarrow{p^2\ll m^2}\;
    24\pi m^4\int_\Sigma\Lambda\,\Box_h^{-1}\Lambda\,\mathrm{vol}_h,
    \label{eq:dual_IR}
\end{equation}
so that $\Lambda$ is gapped in the infrared, consistent with the original
theory being massive.\\

\noindent
At first order in $m^2$, the insertion of the potential
$-\frac{m^2}{8\pi}\int(e^\phi-1)^2\,\mathrm{vol}_h$ into the Liouville path
integral produces a correction to \eqref{eq:dual_nonlocal_forms}. This
correction involves the two-point function
\begin{equation}
    \left\langle(e^{\phi(x)}-1)(e^{\phi(y)}-1)\right\rangle_{L, J}
    \label{eq:two_point_liouville}
\end{equation}
in the Liouville theory with source $J[\Lambda] =
(\mathrm{d}\star_h\mathrm{d}+m^2)\Lambda$. This expectation value depends
non-locally on $\Lambda$ through the Liouville propagator, and generates an
additional non-local interaction for $\Lambda$. In the local approximation where $\Lambda$ is slowly varying, the leading correction is
\begin{equation}
    S_{\mathrm{dual}}^{(1)}[\Lambda]
    = -\frac{m^2}{16\pi}\int_\Sigma
    \left(e^{G_L\star J(x)} - 1\right)^2\mathrm{vol}_h
    + \mathcal{O}(m^4),
    \label{eq:dual_first_order}
\end{equation}
where
\begin{eqnarray}
    G_L\star J(x) = 48\pi\int_\Sigma G_\eta(x,y)(\Box_\eta - m^2)\Lambda(y) \,\mathrm{vol}_h(y)\,,
\end{eqnarray}
is the convolution of $J[\Lambda]$ with the massless
Liouville propagator $G_\eta(x,y) = -\frac{1}{4\pi}\log|x-y|^2$.
For slowly
varying $\Lambda$, $G_L\star J \approx -48\pi\Lambda$ and
\begin{equation}
    S_{\mathrm{dual}}^{(1)}[\Lambda]
    \approx -\frac{m^2}{16\pi}\int_\Sigma
    \left(e^{-48\pi\Lambda} - 1\right)^2\mathrm{vol}_h
    + \mathcal{O}(m^4).
    \label{eq:dual_first_local}
\end{equation}
This is a non-trivial potential for $\Lambda$. Expanding it in small $\Lambda$,
$(e^{-48\pi\Lambda}-1)^2 = (48\pi)^2\Lambda^2 - (48\pi)^3\Lambda^3
+ \frac{7(48\pi)^4}{12}\Lambda^4 + \ldots$\ gives an infinite series of
polynomial interactions for $\Lambda$, beginning with a mass term at order $\Lambda^2$.\\

\noindent
The full dual theory for $\Lambda$, to first order in $m^2$ and in the local
approximation, is therefore
\begin{equation}
    S_{\mathrm{dual}}[\Lambda]
    = S_{\mathrm{dual}}^{(0)}[\Lambda]
    - \frac{m^2}{16\pi}\int_\Sigma
    \left(e^{-48\pi\Lambda} - 1\right)^2\mathrm{vol}_h
    + \mathcal{O}(m^4),
    \label{eq:dual_full}
\end{equation}
where $S_{\mathrm{dual}}^{(0)}$ is the non-local kinetic term
\eqref{eq:dual_nonlocal_forms}. The potential $(e^{-48\pi\Lambda}-1)^2$ is
not a standard sigma-model Kähler potential, nor a sine-Gordon potential, nor
a simple mass term. It is an exponential-squared potential for $\Lambda$,
qualitatively similar in structure to the Toda-field potential but with a
different argument. It is bounded below (the minimum is at $\Lambda = 0$),
consistent with the stability of the original massive scalar theory.\\

\noindent
The central conclusion though, is that \emph{self-duality is broken} 
by the mass.
The dual of the massive scalar under spacetime duality is a non-local theory
with a non-Gaussian potential of the form $(e^{-48\pi\Lambda}-1)^2$. This
theory is neither a massive scalar nor any other standard local QFT. 
It is, as far as we are aware, a
genuinely new theory generated by the spacetime duality transformation, whose
non-locality and non-polynomial potential both reflect the breaking 
of conformal
invariance by the mass. In the geometric language of
Section~\ref{subsec:geometry_review}, the Legendre transform on
$\mathrm{Met}(\Sigma)/\mathrm{Diff}(\Sigma)$ does not preserve the deformed
bundle $\mathcal{L}_m$. Instead, it maps $\mathcal{L}_m$ to a distinct bundle
$\widetilde{\mathcal{L}}_m$ whose sections are the non-local, non-polynomial
functionals of $\Lambda$ described above.

\subsection{The interacting scalar}
\label{subsec:interacting_scalar}

Now let us consider a scalar with a quartic self-interaction, 
linearised via the
Hubbard--Stratonovich identity
\begin{equation}
    -\lambda\int_\Sigma X^4\,\mathrm{vol}_h
    \;\longrightarrow\;
    \int_\Sigma\left(4\lambda X^2\chi - 4\lambda\chi^2\right)\mathrm{vol}_h,
    \label{eq:HS}
\end{equation}
so that integrating out $\chi$ via its algebraic equation of motion
$\chi = X^2/2$ reproduces the original quartic interaction. At fixed $\chi$,
the action is that of a scalar with field-dependent mass $m^2(\chi) =
-8\lambda\chi$. Performing the spacetime duality at fixed $\chi$ via the
analysis of Sections~\ref{subsec:massive_anomaly}--\ref{subsec:scalar_dual}
yields the dual action $S_{\mathrm{dual}}[\Lambda; m^2(\chi)]$ of the form
\eqref{eq:dual_full} with $m^2$ replaced by $m^2(\chi) = -8\lambda\chi$. The
full partition function is then
\begin{equation}
    Z_{\mathrm{dual}} = \int[D\chi]\int[D\Lambda]_h\,
    \exp\!\left(iS_{\mathrm{dual}}[\Lambda; -8\lambda\chi]
    - 4i\lambda\int_\Sigma\chi^2\,\mathrm{vol}_h\right).
    \label{eq:dual_interacting_full}
\end{equation}
Since $\chi$ carries no kinetic term, the path integral over $\chi$ at each
point $x\in\Sigma$ is a local integral. At leading order in $m^2(\chi)$ and in
the local approximation for the dual potential, the $\chi$-dependent part of
the exponent in \eqref{eq:dual_interacting_full} is, from
\eqref{eq:dual_full},
\begin{align}
    &iS_{\mathrm{dual}}[\Lambda; -8\lambda\chi]
    - 4i\lambda\int_\Sigma\chi^2\,\mathrm{vol}_h
    \nonumber\\
    &\quad = iS_{\mathrm{dual}}^{(0)}[\Lambda; m=0]
    + \frac{8i\lambda}
    {16\pi}\int_\Sigma\chi(e^{-48\pi\Lambda}-1)^2\,\mathrm{vol}_h
    - 4i\lambda\int_\Sigma\chi^2\,\mathrm{vol}_h
    + \mathcal{O}(\lambda^2\chi^2),
    \label{eq:chi_exponent}
\end{align}
where $S_{\mathrm{dual}}^{(0)}[\Lambda; m=0]$ is the $m\to 0$ limit of
\eqref{eq:dual_nonlocal_forms}, which reduces to the massless scalar action
$48\pi\int\mathrm{d}\Lambda\wedge\star_h\mathrm{d}\Lambda$. The remaining
$\chi$-dependent terms are quadratic in $\chi$ at each point,
\begin{equation}
    \frac{i\lambda}{2\pi}\int_\Sigma\chi(e^{-48\pi\Lambda}-1)^2\,\mathrm{vol}_h
    - 4i\lambda\int_\Sigma\chi^2\,\mathrm{vol}_h.
    \label{eq:chi_quadratic}
\end{equation}
The $\chi$ path integral at each point is a Gaussian with linear and quadratic
terms,
\begin{eqnarray}
    \int_{-\infty}^\infty &d\chi(x)&\,\exp\!\left(
    \frac{i\lambda\chi(x)}{2\pi}(e^{-48\pi\Lambda(x)}-1)^2
    - 4i\lambda\chi(x)^2
    \right)\mathrm{vol}_h(x)\nonumber\\
    &\propto& \exp\!\left(
    \frac{i\lambda}{64\pi^2}(e^{-48\pi\Lambda(x)}-1)^4
    \right)\mathrm{vol}_h(x),
    \label{eq:chi_gaussian}
\end{eqnarray}
after completing the square in $\chi$ with saddle point
$\chi^*(x) = \frac{1}{16\pi}(e^{-48\pi\Lambda(x)}-1)^2$
and substituting back. The overall constant from the Gaussian
normalisation is $\phi$-independent and is absorbed into the path-integral
measure. The dual partition function after integrating out $\chi$ is therefore
\begin{equation}
    Z_{\mathrm{dual}} = \int[D\Lambda]_h\,\exp\!\left(
    48i\pi\int_\Sigma\mathrm{d}\Lambda\wedge\star_h\mathrm{d}\Lambda
    + \frac{i\lambda}{64\pi^2}\int_\Sigma(e^{-48\pi\Lambda}-1)^4\,\mathrm{vol}_h
    + \mathcal{O}(\lambda^2)
    \right).
    \label{eq:dual_interacting_result}
\end{equation}
The dual theory for the interacting scalar has a free massless kinetic term
for $\Lambda$ and an interaction potential of the form $(e^{-48\pi\Lambda}-1)^4$,
which is an exponential-quartic potential. For small $\Lambda$ this expands as
$(48\pi)^4\Lambda^4 + \ldots$, giving a $\Lambda^4$ interaction consistent with the
original $X^4$ theory. For large $\Lambda$ it saturates to a constant. The
potential is bounded below (minimum at $\Lambda = 0$, value $0$) and above
(maximum value $1$), making the dual theory well-defined.\\

\noindent
Several features of \eqref{eq:dual_interacting_result} deserve comment.
First, the potential $(e^{-48\pi\Lambda}-1)^4$ is \emph{not} a sigma-model
Kähler potential which would give a metric on field space
(a two-derivative term) rather than a potential (a zero-derivative term). The
dual theory \eqref{eq:dual_interacting_result} has a standard kinetic term
and a non-polynomial potential, which is the structure of an ordinary scalar
field theory, albeit with a non-standard potential. Second, the potential is
bounded above, which means that for large $|\Lambda|$ the interaction is
irrelevant and the theory flows to the free massless scalar in the UV, as
expected from dimensional analysis. Third, the coupling constant in the dual
potential is $\lambda/64\pi^2$, loop-suppressed relative to $\lambda$ by a factor $1/64\pi^2$, consistent with the interpretation that the dual
potential arises from integrating out $\chi$, which plays the role of a loop fluctuation in the auxiliary-field language.\\

\noindent
It is worth pausing to appreciate that the result \eqref{eq:dual_interacting_result} is valid only
to leading order in $\lambda$ and in the local (slowly varying $\Lambda$)
approximation. At higher orders in $\lambda$, the $\chi$ integral should produce further corrections to the potential and the full resummation would require
solving the $\chi$ path integral exactly, which should, in turn, generate an all-orders exponential-polynomial potential for $\Lambda$.

%%%%%%%%%%%%%%%%%%%%%%%%%%%%%%%%%%
\section{The Massive Fermion}
\label{sec:massive_fermion}

Having established the spacetime-duality construction for a massive scalar
field, we now turn to the corresponding problem for a massive Dirac fermion.
Although the overall strategy closely parallels the scalar case, the fermionic
theory exhibits two important differences. First, the kinetic term of a
massless Dirac fermion is classically Weyl invariant in two dimensions once
the spinor is assigned its canonical conformal weight. Second, the fermion
mass operator carries a non-trivial Weyl weight and consequently couples to
the conformal mode differently from its scalar counterpart. Determining this
gravitational dressing is the essential first step before discussing the
bosonised description and the corresponding spacetime dual.\\

\noindent
Throughout this section we consider a two-dimensional Lorentzian spin manifold
$(\Sigma,g)$ equipped with zweibein $e^a$, spin connection
$\omega^{ab}$ and associated Dirac operator
\begin{eqnarray}
    \slashed D_g=\gamma^\mu D_\mu,
\qquad
D_\mu
=
\partial_\mu
+
\frac18
\omega_\mu{}^{ab}
[\gamma_a,\gamma_b]\,.
\end{eqnarray}
The action of a massive Dirac fermion is
\begin{equation}
S_m[g,\chi,\bar\chi]
=
-
\int_\Sigma
\bar\chi
\left(
i\slashed D_g+m
\right)
\chi\,
\mathrm{vol}_g,
\label{eq:fermion_action}
\end{equation}
where, as before, $\mathrm{vol}_g$ denotes the metric volume two-form. Also as before, we write the dynamical metric in conformal gauge,
\begin{equation}
g
=
e^\phi h
=
e^{2\sigma}h,
\qquad
\sigma=\frac{\phi}{2},
\label{eq:metric_split}
\end{equation}
with $h$ a fixed reference metric. The corresponding volume form satisfies
$\mathrm{vol}_g = e^\phi \mathrm{vol}_h$. Our first task will be to determine how the conformal mode modifies the fermion mass.

%%%%%%%%%%%%%%%%%%%%%%%%%%%%%%%%%%
\subsection{Gravitational mass dressing}

The simplest derivation proceeds directly from the action. Under a Weyl transformation the Dirac spinor carries conformal weight
$\frac12$, so 
\begin{equation}
\chi
=
e^{-\phi/4}\psi,
\qquad
\bar\chi
=
e^{-\phi/4}\bar\psi.
\label{eq:spinor_weyl}
\end{equation}
This field redefinition is precisely the one that renders the massless Dirac
action independent of the conformal factor,
\begin{equation}
\bar\chi\,i\slashed D_g\chi\,
\mathrm{vol}_g
=
\bar\psi\,i\slashed D_h\psi\,
\mathrm{vol}_h,
\label{eq:massless_weyl}
\end{equation}
reflecting the classical conformal invariance of the two-dimensional massless
Dirac theory. The mass term behaves differently. Substituting
\eqref{eq:spinor_weyl} into
\eqref{eq:fermion_action} immediately gives
\begin{align}
m\,
\bar\chi\chi\,
\mathrm{vol}_g
&=
m
\left(
e^{-\phi/4}\bar\psi
\right)
\left(
e^{-\phi/4}\psi
\right)
e^\phi
\mathrm{vol}_h
\nonumber\\
&=
m
e^{\phi/2}
\bar\psi\psi
\,
\mathrm{vol}_h.
\label{eq:mass_scaling}
\end{align}
The conformal mode therefore dresses the fermion mass according to $m\longrightarrow
m\,e^{\phi/2}$. The origin of the exponent is now transparent. The volume form contributes one
power of $e^\phi$, while the two spinor fields together contribute the
compensating factor $e^{-\phi/2}$, leaving a net factor
$e^{\phi/2}$. Unlike the scalar field, whose Weyl weight vanishes in two
dimensions, the fermion therefore absorbs half of the conformal scaling.\\

\noindent
This fermion mass dressing is the central result of this subsection. It is obtained directly from the
action and therefore does not rely on any properties of the Dirac operator. Nevertheless, it is instructive to verify that exactly the same conclusion
follows from the Weyl covariance of the operator itself.
The Dirac operator is not Weyl invariant but rather Weyl covariant.
Under the standard transformation of the spin connection\footnote{The torsion-free condition  $\mathrm{d}\tilde{e}^a
+ \tilde\omega^a{}_b\wedge\tilde{e}^b = 0$ for the
rescaled zweibein $\tilde{e}^a = e^\sigma e^a$ determines
the transformed spin connection. A direct computation gives
\begin{equation}
    \tilde\omega^{ab}
    =
    \omega^{ab}
    +
   e^a\wedge\iota^b\mathrm{d}\sigma
    -
    e^b\wedge\iota^a\mathrm{d}\sigma,
    \label{eq:spin_connection_forms}
\end{equation}
where $\iota^a = e^{a\mu}\iota_\mu$ denotes the interior product with
the dual frame vector $e^{a\mu}$, so that
$\iota^a\mathrm{d}\sigma = e^{a\mu}\partial_\mu\sigma$.
In components, \eqref{eq:spin_connection_forms} reads
\begin{equation}
    \tilde\omega^{ab}_\mu
    =
    \omega^{ab}_\mu
    +
    e^a_\mu\,e^{b\nu}\partial_\nu\sigma
    -
    e^b_\mu\,e^{a\nu}\partial_\nu\sigma,
    \label{eq:spin_connection_components}
\end{equation}
which is manifestly antisymmetric in $a,b$ as required.},
a direct contraction gives
\begin{equation}
i\slashed{D}_{e^{2\sigma}h}
=
e^{-\sigma}
\!\left(
i\slashed{D}_h
+
\frac{i}{2}\slashed{d}\sigma
\right),
\label{eq:dirac_weyl_additive}
\end{equation}
where $\slashed{d}\sigma = \gamma^\mu\partial_\mu\sigma$.
Acting with $i\slashed{D}_g + m$ on $\chi = e^{-\sigma/2}\psi$
and using \eqref{eq:dirac_weyl_additive},
\begin{align}
i\slashed{D}_g\!\left(e^{-\sigma/2}\psi\right)
&=
e^{-\sigma}
\!\left(
i\slashed{D}_h + \frac{i}{2}\slashed{d}\sigma
\right)\!
\left(e^{-\sigma/2}\psi\right)
\nonumber\\
&=
e^{-3\sigma/2}
\!\left(
i\slashed{D}_h\psi
-
\frac{i}{2}\slashed{d}\sigma\,\psi
+
\frac{i}{2}\slashed{d}\sigma\,\psi
\right)
\nonumber\\
&=
e^{-3\sigma/2}\,
i\slashed{D}_h\,\psi.
\label{eq:dirac_expansion}
\end{align}
Adding the mass term $m\chi = me^{-\sigma/2}\psi = e^{-3\sigma/2}me^\sigma\psi$
therefore yields
\begin{equation}
\left(i\slashed{D}_g + m\right)\chi
=
e^{-3\sigma/2}
\left(
i\slashed{D}_h
+
m\,e^{\sigma}
\right)\psi,
\label{eq:weyl_dirac_operator}
\end{equation}
This operator-level derivation then reproduces precisely the same mass dressing as
the action-level calculation,
where $m
\longrightarrow
m e^\sigma
=
m e^{\phi/2}$,
providing a non-trivial consistency check. The agreement of the two derivations establishes that the universal gravitational dressing of the fermion mass is
\begin{equation}
m
\longrightarrow
m e^{\phi/2},
\label{eq:fermion-mass-dressing}
\end{equation}
and this will play the same role in the fermionic theory as
$m^2e^\phi$ did in the scalar theory. Having now established that the only effect of the Weyl
transformation on the massive Dirac action is the universal replacement \eqref{eq:fermion-mass-dressing}, obtained directly from the fermionic action and  therefore
independent of any bosonisation procedure, we now ask how this dressing is reflected in the bosonised description.\\

\noindent
On a fixed background, the Coleman--Mandelstam correspondence identifies the
fermion mass operator with the lowest-dimensional neutral vertex operator of
the compact boson,
\begin{equation}
\bar\psi\psi
\;\longleftrightarrow\;
C\,:\!\cos(\beta\vartheta)\!:,
\label{eq:mass_dictionary}
\end{equation}
where $C$ is a scheme-dependent normalisation constant and $\vartheta$ denotes
the compact boson. Since the fermion mass appears linearly in the
action, the Weyl transformation
\eqref{eq:fermion-mass-dressing} immediately suggests that the bosonised interaction acquires the same multiplicative dressing,
\begin{equation}
\mu
\cos(\beta\vartheta)
\quad\longrightarrow\quad
\mu\,e^{\phi/2}\cos(\beta\vartheta),
\label{eq:dressed_vertex}
\end{equation}
where $\mu\propto m$ is the renormalised sine-Gordon coupling.
Equation \eqref{eq:dressed_vertex} has a simple geometric interpretation.
Unlike the compact scalar $\vartheta$, which parametrises the bosonised matter
sector, the field $\phi$ is the conformal mode of the spacetime metric. Consequently the interaction does not modify the periodicity of the
sine-Gordon potential. Rather, the geometry rescales the local strength of the
interaction through the Weyl factor $e^{\phi/2}$ while leaving its periodic
structure intact. The conformal mode therefore acts as a position-dependent
coupling for the vertex operator.\\

\noindent
Equivalently, the dressed interaction may be written as
\begin{equation}
\mu e^{\phi/2}\cos(\beta\vartheta)
=
\frac{\mu}{2}
\left(
e^{\phi/2}V_+
+
e^{\phi/2}V_-
\right)\,
\label{eq:dressed_vertices}
\end{equation}
in terms of the vertex operators
$V_\pm = :\!e^{\pm i\beta\vartheta}\!:$. Each chiral vertex is therefore multiplied by the same exponential of the
Liouville field. In this sense the conformal mode dresses the bosonised mass operator exactly as it dresses the fermion mass in the original Dirac theory.\\

\noindent
It is important to distinguish this dressing from the standard Liouville gravitational dressing familiar from two-dimensional quantum gravity. Equation \eqref{eq:dressed_vertex} is not obtained by demanding conformal
invariance of the vertex operator or by solving the KPZ scaling relations. Instead, it follows from the much simpler observation that the fermion mass
appears only as a coefficient multiplying the operator
$\bar\psi\psi$. Since the Weyl transformation rescales this coefficient by
$e^{\phi/2}$, the same factor necessarily accompanies its bosonised image. This argument should therefore be regarded as an action-level derivation of
the gravitational dressing rather than a derivation of curved-space
bosonisation itself. A complete formulation of bosonisation on an arbitrary
curved background would require a careful treatment of the conformal anomaly, normal ordering of composite operators and the renormalisation of the
functional measure. Those questions lie beyond the scope of the present work. For our purposes, the universal dressing
\eqref{eq:dressed_vertex} is sufficient to determine how the conformal mode enters the bosonised theory.\\

\noindent
The resulting bosonic action is
\begin{equation}
S_{\rm bos}[h,\phi,\vartheta]
=
\frac{1}{2}
\int_\Sigma
\mathrm{d}\vartheta
\wedge
\star_h
\mathrm{d}\vartheta
+
\mu
\int_\Sigma
e^{\phi/2}
\cos(\beta\vartheta)\,
\mathrm{vol}_h,
\label{eq:dressed_SG}
\end{equation}
where the kinetic term is written with respect to the fixed reference metric
$h$, while the interaction is dressed by the conformal mode. The appearance
of the product
\begin{equation}
e^{\phi/2}\cos(\beta\vartheta)
\end{equation}
is the central result of the fermionic construction. It shows that the
spacetime geometry couples multiplicatively to the sine-Gordon interaction,
providing the fermionic analogue of the exponential potential that arose in
the massive scalar theory.

%%%%%%%%%%%%%%%%%%%%%%%%%%%%%%%%%%
\subsection{Towards the Dual Theory}
\label{subsec:fermion_dual}

For the massive scalar field, the corresponding stage of the construction consisted of integrating over the conformal mode to obtain an effective theory
for the Lagrange multiplier field $\Lambda$. The resulting functional integral
was governed by a single scalar degree of freedom and led directly to the dual
description discussed in Section~\ref{sec:massive_scalar}.\\

\noindent
The fermionic theory differs in one important respect. After bosonisation, the
matter sector is no longer described solely by the conformal mode but by the
pair of scalars
$(\phi,\vartheta)$, where $\phi$ parametrises the geometry through the Weyl factor and
$\vartheta$ is the compact boson replacing the fermionic degrees of freedom. The natural starting point for the spacetime-duality construction is therefore
the coupled action
\begin{equation}
S[\phi,\vartheta,\Lambda]
=
S_{\rm grav}[\phi,\Lambda]
+
\frac{1}{2}
\int_\Sigma
d\vartheta
\wedge
\star_h
d\vartheta
+
\mu
\int_\Sigma
e^{\phi/2}
\cos(\beta\vartheta)\,
\mathrm{vol}_h,
\label{eq:coupled_action}
\end{equation}
where $S_{\rm grav}[\phi,\Lambda]$ denotes the gravitational sector obtained
earlier, including the Liouville action, the cosmological
constant and the Lagrange-multiplier constraint. Unlike the scalar theory, the conformal mode now appears only through the
interaction
\begin{equation}
e^{\phi/2}\cos(\beta\vartheta),
\label{eq:interaction}
\end{equation}
which couples the geometric and matter degrees of freedom
multiplicatively. Consequently neither $\phi$ nor $\vartheta$ can be
integrated out independently. Instead, the functional integral becomes
\begin{equation}
Z
=
\int
[D\phi]
[D\vartheta]
[D\Lambda]\,
e^{\,iS[\phi,\vartheta,\Lambda]},
\label{eq:coupled_partition}
\end{equation}
representing a genuinely interacting two-field theory.
From the perspective of spacetime duality, the problem is therefore no longer one of integrating out a single conformal mode but of understanding the joint
dynamics of the Liouville field and the compact boson. This interaction is
reminiscent of Liouville-dressed vertex operators in two-dimensional gravity,
although its origin here is rather different. In the present construction the
dressing is inherited directly from the Weyl transformation of the massive
Dirac action and is therefore fixed at the classical level before any
consideration of conformal invariance or KPZ scaling.\\

\noindent
Several possible strategies suggest themselves. We could integrate over the
compact boson first, obtaining an effective action for the conformal mode and
the Lagrange multiplier. Alternatively, we may regard the interaction
\eqref{eq:interaction} as a perturbation of Liouville theory and develop a
systematic expansion in the dressed vertex operators
$e^{\phi/2}e^{\pm i\beta\vartheta}$. A third possibility is to study the coupled renormalisation-group flow of the two scalar fields and determine whether the spacetime-duality transformation
admits a natural description in terms of the resulting infrared fixed point. Each of these approaches lies beyond the scope of the present work.\\

\noindent
The essential conclusion of this section is therefore that the fermionic
construction naturally leads not to an isolated sine-Gordon theory but to a
coupled Liouville--sine-Gordon system,
\begin{equation}
S_{\rm LSG}
=
S_{\rm grav}[\phi,\Lambda]
+
\frac{1}{2}
\int_\Sigma
d\vartheta
\wedge
\star_h
d\vartheta
+
\mu
\int_\Sigma
e^{\phi/2}
\cos(\beta\vartheta)\,
\mathrm{vol}_h.
\label{eq:LSG_action}
\end{equation}
The appearance of this interaction is a direct consequence of the universal
Weyl dressing \eqref{eq:fermion-mass-dressing}, together with the standard bosonisation of the fermion mass operator. It
provides the natural fermionic analogue of the scalar construction developed
in Section~\ref{sec:massive_scalar} and identifies the coupled Liouville--sine-Gordon theory
as the appropriate starting point for a complete spacetime-duality programme
for massive Dirac fermions. A detailed analysis of the functional integral
\eqref{eq:coupled_partition}, together with the corresponding dual theory for
the Lagrange multiplier field $\Lambda$, will be presented elsewhere.

%%%%%%%%%%%%%%%%%%%%%%%%%%%%%%%%%%
\section{Discussion}
\label{sec:discussion}

The central organising principle behind the spacetime duality construction of \cite{BQ1998} is the gauging of a spacetime symmetry by coupling the matter theory to a dynamical metric,
and then imposing a constraint via a Lagrange multiplier field $\Lambda$ that
forces the dynamical metric to coincide with a fixed background. The dual theory is obtained by exchanging the order of functional integration. In the
massless case this procedure is under full control since the path-integral measure
of a CFT of central charge $c$ is a section of the determinant line bundle $\mathcal{L} \to \mathrm{Met}(\Sigma)/\mathrm{Diff}(\Sigma)$ equipped with
the Quillen metric \cite{Quillen1985}, and the Liouville action is the
log-norm of this section along a conformal orbit, with curvature fixed by the
Belavin--Knizhnik theorem \cite{BelavinKnizhnik1986} to be $c_1(\mathcal{L})
= (c/12)\,\omega_{WP}$. The spacetime duality transformation is then a
fibre-direction Legendre transform on $\mathcal{L}$. Self-duality of the
massless scalar corresponds to this transform being an automorphism of
$\mathcal{L}$, while bosonisation of the massless fermion corresponds to an
isomorphism $\mathcal{L}_{\mathrm{Dirac}} \cong \mathcal{L}_{\mathrm{scalar}}$.
This is really the geometric content of the determinant identity \eqref{eq:det_identity_forms}.\\

\noindent
The main aim of this paper was to extend this construction to massive theories. A massive field theory is of course not a CFT, so its path-integral measure is not a
section of $\mathcal{L}$ in the standard sense. The natural replacement is a
deformed bundle $\mathcal{L}_m$, whose curvature receives corrections in $m^2$
via the full heat kernel. A naive expectation is that the leading correction
is a cosmological-constant term $\sim m^2(e^\phi-1)$ in the conformal-mode
direction. Our central technical finding of Section~\ref{sec:massive_scalar}
is that this expectation is wrong. Tracking three distinct contributions (the path-integral measure anomaly, the
matter determinant, and the $\Delta_{LM}$ constraint prefactor) to
the conformal-mode effective action, we showed
that the naive leading correction cancels exactly between the three sources,
as required by the internal consistency of the master path integral. The
genuine leading deformation of $\mathcal{L}_m$ first appears at second order
in $(e^\phi-1)$, arising from the $n=2$ term in the resolvent expansion of
the massive propagator via the spatial integral 
\begin{eqnarray}
    \int G_m(x,y)^2\,\mathrm{vol}_h(y)
    = \frac{1}{4\pi m^2}\,.
\end{eqnarray}  
The resulting curvature correction
$-\frac{m^2}{16\pi}\int_\Sigma(e^\phi-1)^2\,\mathrm{vol}_h$ is not captured
by any finite order of the local Seeley--DeWitt expansion; it is intrinsically
a non-local, two-point effect of the massive propagator. This is the geometric
source of the breaking of self-duality. The $(e^\phi-1)^2$ potential renders
the conformal-mode path integral non-Gaussian, so the Legendre transform no
longer closes in a form that reproduces the original action. The dual theory
for $\Lambda$ is correspondingly non-local, with a kinetic operator
interpolating between the massless free-scalar result in the ultraviolet and
a gapped theory in the infrared, and does not correspond to a standard CFT
section of $\mathcal{L}$ or $\mathcal{L}_m$.\\

\noindent
For the massive fermion the situation is geometrically richer. Through two independent
derivations, we established that the fermion mass dresses under the conformal rescaling $g = e^\phi h$ as $m \to m\,e^{\phi/2}$. The
exponent $\phi/2$, rather than $\phi$ as for the scalar, reflects the Weyl
weight $1/2$ of the two-dimensional Dirac spinor and is the fundamental
geometric distinction between the scalar and fermionic constructions. Via the
standard Coleman--Mandelstam bosonisation \cite{Coleman1975, Mandelstam1975},
this dressing transfers to the mass bilinear as
$\mu\cos(\beta\vartheta) \to \mu\,e^{\phi/2}\cos(\beta\vartheta)$, where
$\vartheta$ is the compact boson dual to the fermion and $\beta^2 = 4\pi$
at the free-fermion point. The resulting matter sector is a coupled Liouville--sine-Gordon system
\eqref{eq:LSG_action}, in which the conformal mode $\phi$ and the bosonised
field $\vartheta$ interact multiplicatively through $e^{\phi/2}\cos(\beta
\vartheta)$. We have identified this system as the appropriate starting point
for the fermionic spacetime-duality programme but have not completed the
construction. Deriving the dual action for $\Lambda$ from the functional
integral \eqref{eq:coupled_partition} over the coupled
$(\phi,\vartheta,\Lambda)$ system is the natural next step and is left for
future work.\\ 

\noindent
The factorisation $\mathrm{Met}(\Sigma)/\mathrm{Diff}(\Sigma) \cong
\mathrm{Conf}(\Sigma) \times \mathcal{T}_g$ raises a question about the role
of the Teichm\"{u}ller moduli $\mathcal{T}_g$ that neither the original construction in \cite{BQ1998} nor the present work addresses. The constraint $\Delta_{LM}$ pins
the conformal factor within a fixed conformal class but is silent about
$\mathcal{T}_g$. On a surface of non-trivial topology, a complete treatment
would need to specify how $\Delta_{LM}$ acts on the moduli; in particular,
whether the conformal mode should be treated as a circle-valued field $\phi
\sim \phi + 2\pi$ satisfying a Dirac quantisation condition, so that the path
integral includes a sum over winding sectors. Such a sum would produce a
theta-function structure analogous to Witten's sum over winding numbers of
the axion field in the wormhole geometry \cite{Witten2026}. There, the duality
between a massless scalar and a two-form gauge field in four Euclidean dimensions requires a Poisson resummation over flux sectors rather than a
classical field map, with the partition function in the scalar description decomposing as a sum over winding numbers that Poisson-resums into a sum over
integer fluxes of the dual two-form. If the conformal mode were
circle-valued with Dirac-quantised periods, an analogous resummation would
generate a periodic potential for $\Lambda$, making the analogy between the two constructions exact. We regard this as one of the most interesting open questions raised by the present work; the physical and
geometric interpretation of a circle-valued conformal mode within the
spacetime symmetry gauging framework, and whether imposing such a quantisation
condition changes the massless self-duality result or the massive results
derived here.\\ 

\noindent
Finally, we comment on the prospects for extending spacetime duality to $2+1$
dimensions, which motivated some of the geometric framing developed here. The
entire mechanism in $1+1$ dimensions relies on three special features of two
dimensions: the uniformisation theorem (every metric is conformally flat, so
the metric path integral reduces to a single scalar $\phi$); the identity
$\sqrt{-g}\mathcal{R}_g - \sqrt{-h}\mathcal{R}_h = \sqrt{-h}\,\Box_h\phi$,
which makes the Lagrange-multiplier coupling linear in $\phi$; and the local
Weyl anomaly, which gives the Liouville action as a controlled, local output
of the conformal anomaly. None of these features hold in three dimensions.
The natural route into $2+1$ dimensions is via the Chern--Simons formulation
of three-dimensional gravity \cite{WittenCS1988}, in which the vielbein and
spin connection are packaged into a gauge connection for $ISO(2,1)$, and
non-conformal-flatness of the background metric manifests as non-trivial
holonomy of the flat connection rather than as local tensor degrees of freedom.
In this formulation, gauging Lorentz invariance reduces to gauging the
$ISO(2,1)$ internal symmetry of the Chern--Simons theory, which suggests that
the spacetime duality programme in $2+1$ dimensions is more naturally framed
as a problem within the three-dimensional duality web
\cite{Seiberg2016, WittenCS1988} than as a direct generalisation of the
two-dimensional construction. Whether this leads to genuinely new results or, as seems likely for the conformally flat sector, reproduces known dualities
with a new derivation, is a question we leave open.

%%%%%%%%%%%%%%%%%%%%%%%%%%%%%%%%%%
\section*{Acknowledgements}
We would like to thank Fernando Quevedo and Edward Witten for several useful and clarifying discussions over the (rather long) time that we have been thinking about this problem. JM is supported in part by the ``Quantum Technologies for Sustainable Development" grant
from the National Institute for Theoretical and Computational Sciences of South Africa
(NITheCS) and would like to thank the ICTP-SAIFR for their kind hospitality while this work was being completed. The work of HN is supported in part by  CNPq grant 304583/2023-5 and FAPESP grant 2024/15298-0. HN would also like to thank the ICTP-SAIFR for their support  through FAPESP grant 2021/14335-0.

%%%%%%%%%%%%%%%%%%%%%%%%%%%%%%%%%%
\appendix 
\section{The Seeley--DeWitt Expansion}
\label{app:seeley_dewitt}

The Seeley--DeWitt expansion is an asymptotic small-$s$ expansion of the heat
kernel of an elliptic differential operator on a Riemannian manifold. It
underlies the heat-kernel computations of Section~\ref{sec:massive_scalar} and
we collect here the relevant definitions, general formulae, and a worked
example.

\subsection{The heat kernel and heat trace}
\label{app:heat_kernel_def}

Let $(\mathscr{M}, g)$ be a smooth Riemannian manifold of dimension $n$,
without boundary, and let
\begin{equation}
    \mathcal{O} = -\Box_g + E
    \label{eq:general_operator}
\end{equation}
be a second-order elliptic operator, where $\Box_g = g^{\mu\nu}\nabla_\mu
\nabla_\nu$ is the scalar Laplacian and $E \in C^\infty(\mathscr{M})$ is a
smooth potential. The \emph{heat kernel} $K(s, x, y)$ is the fundamental
solution of the heat equation
\begin{equation}
    \left(\partial_s + \mathcal{O}_x\right)K(s, x, y) = 0,
    \qquad
    \lim_{s \to 0^+} K(s, x, y) = \frac{\delta^{(n)}(x - y)}{\sqrt{g(x)}},
    \label{eq:heat_equation}
\end{equation}
where the subscript $x$ on $\mathcal{O}_x$ indicates that the operator acts on
the first argument. The \emph{heat trace} is the integral of the coincident
heat kernel,
\begin{equation}
    \mathrm{tr}\!\left[e^{-s\mathcal{O}}\right]
    = \int_\mathcal{M} K(s, x, x)\,\mathrm{vol}_g(x).
    \label{eq:heat_trace}
\end{equation}
The heat trace encodes the spectral information of $\mathcal{O}$. Specifically, if
$\{\lambda_k\}$ are the eigenvalues of $\mathcal{O}$, then
$\mathrm{tr}[e^{-s\mathcal{O}}] = \sum_k e^{-s\lambda_k}$. It is related to
the functional determinant by the Schwinger proper-time representation
\begin{equation}
    \log\det\mathcal{O}
    = -\int_0^\infty \frac{\mathrm{d}s}{s}\,
    \mathrm{tr}\!\left[e^{-s\mathcal{O}}\right],
    \label{eq:schwinger}
\end{equation}
which requires UV regularisation near $s = 0$ and IR regularisation near
$s = \infty$. For operators of the form $\mathcal{O} = -\Box_g + m^2 + E$ with
$m^2 > 0$, the IR divergence is absent and we can write
\begin{equation}
    \log\det(-\Box_g + m^2 + E)
    = -\int_0^\infty \frac{\mathrm{d}s}{s}\,e^{-m^2 s}\,
    \mathrm{tr}\!\left[e^{s(\Box_g - E)}\right].
    \label{eq:schwinger_massive}
\end{equation}

\subsection{The asymptotic expansion}
\label{app:asymptotic}

For small $s > 0$, the heat trace has the asymptotic expansion
\begin{equation}
    \mathrm{tr}\!\left[e^{-s\mathcal{O}}\right]
    \sim \frac{1}{(4\pi s)^{n/2}}
    \sum_{k=0}^\infty s^k
    \int_\mathcal{M} a_k(x; \mathcal{O})\,\mathrm{vol}_g(x)
    \qquad \text{as } s \to 0^+,
    \label{eq:SD_expansion}
\end{equation}
where the $a_k(x; \mathcal{O})$ are the \emph{Seeley--DeWitt coefficients}
\cite{Seeley1967, dewitt1965dynamical} and $n$ is the dimension of the 
manifold. The symbol $\sim$ denotes an asymptotic
expansion; the series diverges for any fixed $s > 0$ if carried to all orders,
but the partial sums approximate the heat trace to increasing accuracy as
$s \to 0^+$. On a manifold without boundary, only integer powers of $s$ appear,
so coefficients with non-integer index vanish. The first three non-vanishing
coefficients are
\begin{align}
    a_0(x) &= 1\,,
    \label{eq:a0} \\
    a_1(x) &= \frac{\mathcal{R}}{6} - E\,,
    \label{eq:a1} \\
    a_2(x) &= \frac{1}{180}\!\left(
    \mathcal{R}_{\mu\nu\rho\sigma}\mathcal{R}^{\mu\nu\rho\sigma}
    - \mathcal{R}_{\mu\nu}\mathcal{R}^{\mu\nu}
    + \frac{5}{2}\mathcal{R}^2
    - 6\Box_g\mathcal{R}\right)
    + \frac{\mathcal{R}E}{12}
    - \frac{\Box_g E}{2}
    + \frac{E^2}{2}\,,
    \label{eq:a2}
\end{align}
where $\mathcal{R}$, $\mathcal{R}_{\mu\nu}$, and
$\mathcal{R}_{\mu\nu\rho\sigma}$ are the Ricci scalar, Ricci tensor, and
Riemann tensor of $g$ respectively. Several structural features are worth
noting. First, each $a_k(x)$ is a \emph{local} expression built from the curvature
of $g$ and the potential $E$ at the point $x$; no non-local (integral)
expressions appear. This locality is what makes the Seeley--DeWitt expansion
useful for computing conformal anomalies, since the anomaly must be a local
functional of the metric. Second, $a_0 = 1$ is independent of the geometry
and reflects only the leading-order diffusion on flat space. Third, for a
massless scalar ($E = 0$) on a curved background
\begin{equation}
    a_1(x)\big|_{E=0} = \frac{\mathcal{R}}{6},
    \label{eq:a1_massless}
\end{equation}
which reproduces the standard result that the one-loop conformal anomaly of a
scalar field is proportional to the Ricci scalar. For the massive scalar with a constant negative potential $E = -m^2$,
\begin{equation}
    a_1(x)\big|_{E=-m^2} = \frac{\mathcal{R}}{6} + m^2,
    \label{eq:a1_massive}
\end{equation}
so that
\begin{equation}
    \mathrm{tr}\!\left[e^{-s(-\Box_g + m^2)}\right]
    \sim \frac{e^{-m^2 s}}{4\pi s}
    \int_\mathcal{M}\!\mathrm{vol}_g
    \left(1 + \frac{s\mathcal{R}}{6} + \mathcal{O}(s^2)\right),
    \label{eq:SD_massive_scalar}
\end{equation}
where the factor $e^{-m^2 s}$ arises from the decomposition
$\mathcal{O} = -\Box_g + m^2$ and the convention \eqref{eq:schwinger_massive}.
This is the form that we have used throughout Section~\ref{sec:massive_scalar}.

\subsection{Application to functional determinants}
\label{app:functional_dets}

Substituting \eqref{eq:SD_massive_scalar} into the Schwinger representation
\eqref{eq:schwinger_massive} and integrating term by term gives
\begin{align}
    \log\det(-\Box_g + m^2)
    &= -\frac{1}{4\pi}\int_\mathcal{M}\!\mathrm{vol}_g
    \int_0^\infty \frac{\mathrm{d}s}{s^2}\,e^{-m^2 s}
    \left(1 + \frac{s\mathcal{R}}{6} + \mathcal{O}(s^2)\right)
    \nonumber \\
    &= -\frac{1}{4\pi}\int_\mathcal{M}\!\mathrm{vol}_g
    \left(\int_0^\infty \frac{\mathrm{d}s}{s^2}\,e^{-m^2 s}
    + \frac{\mathcal{R}}{6}\int_0^\infty \frac{\mathrm{d}s}{s}\,e^{-m^2 s}
    + \mathcal{O}(s^0)
    \right).
    \label{eq:logdet_expand}
\end{align}
Both integrals are UV-divergent at $s \to 0$. Introducing a UV cutoff
$s \geq \epsilon$:
\begin{equation}
    \int_\epsilon^\infty \frac{\mathrm{d}s}{s^2}\,e^{-m^2 s}
    = \frac{1}{\epsilon} - m^2\log(m^2\epsilon) + m^2(\gamma_E - 1) + \mathcal{O}(\epsilon),
    \label{eq:div1}
\end{equation}
\begin{equation}
    \int_\epsilon^\infty \frac{\mathrm{d}s}{s}\,e^{-m^2 s}
    = -\log(m^2\epsilon) - \gamma_E + \mathcal{O}(\epsilon),
    \label{eq:div2}
\end{equation}
where $\gamma_E$ is the Euler--Mascheroni constant. The divergence
$1/\epsilon$ in \eqref{eq:div1} is a cosmological-constant counterterm; the
$\log\epsilon$ divergences in both \eqref{eq:div1} and \eqref{eq:div2} are
absorbed into renormalisation of the couplings. After renormalisation, the
finite part of $\log\det(-\Box_g+m^2)$ contains a term
$\frac{m^2}{4\pi}\log m^2\int_\mathcal{M}\mathrm{vol}_g$ and a curvature
correction $\frac{\mathcal{R}}{24\pi}\log m^2\int_\mathcal{M}\mathrm{vol}_g$,
both of which depend on the renormalisation scale. The renormalisation-scale
independent part, which captures the $\phi$-dependence relevant for the
conformal anomaly, is what is computed in Section~\ref{sec:massive_scalar}.

\subsection{Worked example: the Laplacian on \texorpdfstring{$S^2$}{S2}}
\label{app:S2_example}

Let us now illustrate the Seeley--DeWitt expansion 
with the simplest non-trivial
example: the scalar Laplacian $\mathcal{O} = -\Box_{S^2}$ on the round
two-sphere $S^2$ of radius $r$, with no potential ($E = 0$).\\

\noindent
On $S^2$, the volume is $\int_{S^2}\mathrm{vol}_g = 4\pi r^2$ and the Ricci
scalar $\mathcal{R} = 2/r^2$ is of course constant. With $n = 2$ and 
$E = 0$, the
expansion \eqref{eq:SD_expansion} gives
\begin{equation}
    \mathrm{tr}\!\left[e^{s\Box_{S^2}}\right]
    \sim \frac{1}{4\pi s}\cdot 4\pi r^2
    \left(1 + \frac{s}{6}\cdot\frac{2}{r^2} + \mathcal{O}(s^2)\right)
    = \frac{r^2}{s} + \frac{1}{3} + \mathcal{O}(s).
    \label{eq:SD_S2}
\end{equation}
The eigenvalues of $-\Box_{S^2}$ are $\ell(\ell+1)/r^2$ with degeneracy
$2\ell + 1$, so the heat trace is exactly
\begin{equation}
    \mathrm{tr}\!\left[e^{s\Box_{S^2}}\right]
    = \sum_{\ell=0}^\infty (2\ell + 1)\,e^{-s\ell(\ell+1)/r^2}.
    \label{eq:S2_exact}
\end{equation}
For small $s$, this sum is dominated by large $\ell$. Writing $u = \ell + 1/2$
so that $\ell(\ell+1) = u^2 - 1/4$ and $2\ell + 1 = 2u$ we find that
\begin{equation}
    \mathrm{tr}\!\left[e^{s\Box_{S^2}}\right]
    = e^{s/4r^2}\sum_{\ell=0}^\infty 2u\,e^{-su^2/r^2}.
    \label{eq:S2_rewrite}
\end{equation}
Approximating the sum by an integral (valid for $s/r^2 \ll 1$) gives
\begin{equation}
    \sum_{\ell=0}^\infty 2u\,e^{-su^2/r^2}
    \approx \int_0^\infty 2u\,e^{-su^2/r^2}\,\mathrm{d}u
    = \frac{r^2}{s}.
    \label{eq:sum_to_integral}
\end{equation}
Then, expanding $e^{s/4r^2} = 1 + s/4r^2 + \mathcal{O}(s^2)$,
\begin{equation}
    \mathrm{tr}\!\left[e^{s\Box_{S^2}}\right]
    \approx \left(1 + \frac{s}{4r^2} + \mathcal{O}(s^2)\right)\frac{r^2}{s}
    = \frac{r^2}{s} + \frac{1}{4} + \mathcal{O}(s).
    \label{eq:S2_integral_approx}
\end{equation}
The leading term $r^2/s$ agrees between \eqref{eq:SD_S2} and
\eqref{eq:S2_integral_approx}, confirming the $a_0$ coefficient. The constant
terms differ with the Seeley--DeWitt expansion giving $1/3$, while the
integral approximation gives $1/4$. This discrepancy arises because replacing
the sum \eqref{eq:S2_rewrite} by an integral \eqref{eq:sum_to_integral}
neglects the Euler--Maclaurin correction, whose leading term is
\begin{equation}
    \frac{1}{2}\cdot 2u\,e^{-su^2/r^2}\Big|_{u=1/2}
    = e^{-s/4r^2}
    \approx 1 - \frac{s}{4r^2} + \mathcal{O}(s^2).
    \label{eq:EM_correction}
\end{equation}
Adding this correction to \eqref{eq:sum_to_integral} and multiplying by
$e^{s/4r^2}$ gives
\begin{equation}
    e^{s/4r^2}\left(\frac{r^2}{s} + 1\right)
    = \frac{r^2}{s} + 1 + \frac{1}{4} + \mathcal{O}(s)
    = \frac{r^2}{s} + \frac{5}{4} + \mathcal{O}(s),
    \label{eq:over_corrected}
\end{equation}
which overshoots. The correct constant term $1/3$ is obtained only by
resumming the full Euler--Maclaurin series, which is equivalent to using
zeta-function regularisation. Specifically, the exact constant term is
determined by the zeta function of $-\Box_{S^2}$,
\begin{equation}
    \zeta_{S^2}(0) = \mathrm{tr}\!\left[(-\Box_{S^2})^0\right]_{\mathrm{reg}}
    = \frac{1}{3}\,,
    \label{eq:zeta_S2}
\end{equation}
a well-known result \cite{Hawking1977}. The Seeley--DeWitt coefficient
$a_1 = \mathcal{R}/6$ correctly reproduces this since $\frac{1}{4\pi}\cdot
\frac{2/r^2}{6}\cdot 4\pi r^2 = \frac{1}{3}$.\\

\noindent
This $S^2$ example illustrates two points that are important for the
applications in Section~\ref{sec:massive_scalar}. First, the leading
Seeley--DeWitt coefficient $a_0 = 1$ is exact and captures the dominant
small-$s$ behaviour reliably; this is what we use when computing the leading
$(e^\phi - 1)$ term in the conformal anomaly. Second, the subleading
coefficient $a_1$ requires either the full Euler--Maclaurin resummation or
zeta-function methods to extract correctly from a spectral sum; this is why
we treat the $\mathcal{O}(\mathcal{R}/m^2)$ corrections in
Section~\ref{sec:massive_scalar} as subleading and do not compute them
explicitly. The second-order $(e^\phi - 1)^2$ term computed via the resolvent
expansion in Section~\ref{subsec:second_order} is not captured by the
Seeley--DeWitt expansion at any finite order in $s$. It arises instead from
the $n = 2$ term in the resolvent series \eqref{eq:resolvent_expand}, which
involves the spatial integral of $G_m(x,y)^2$ and is genuinely beyond the
local heat-kernel expansion.
%=========================
% Bibliography
%=========================

\bibliographystyle{JHEP}
\bibliography{biblio.bib}

\end{document}